\documentclass[%12pt,%twocolumn,preprint,
preprintnumbers,showpacs,amsmath,amssymb,aps,prd]{revtex4}
\usepackage{dcolumn}
\usepackage{amsthm,amscd,mathrsfs}
\usepackage{amsfonts,bm,latexsym}
\def\be{\begin{equation}}
\def\ee{\end{equation}}
\def\bea{\begin{eqnarray}}
\def\eea{\end{eqnarray}}
\def\ba{\begin{array}}
\def\ea{\end{array}}
\def\nn{\nonumber}
\def\p{\partial}
\def\cD{\mathcal{D}}
\def\cL{\mathcal{L}}
\def\Id{1\!\!1}

\begin{document}
%\begin{CJK*}{GB}{}
\preprint{arXiv:0902.2823v4 [hep-th]}

\title{Separability of a modified Dirac equation in a five-dimensional rotating,
charged black hole in string theory}

\author{Shuang-Qing Wu} % \footnote{Electronic address: sqwu@phy.ccnu.edu.cn}}
%\author{Shuang-Qing Wu (ÎâË«Çå)} % \footnote{Electronic address: sqwu@phy.ccnu.edu.cn}}
\affiliation{College of Physical Science and Technology, Central China Normal University,
Wuhan, Hubei 430079, People's Republic of China}
%\date{\today}

\begin{abstract}
The aim of this paper is to investigate the separability of a spin-$1/2$ spinor field in
a five-dimensional rotating, charged black hole constructed by Cveti\v{c} and Youm in string
theory, in the case when three $U(1)$ charges are set equal. This black hole solution represents
a natural generalization of the famous four-dimensional Kerr-Newman solution to five
dimensions with the inclusion of a Chern-Simons term to the Maxwell equation. It is shown
that the usual Dirac equation can not be separated by variables in this general spacetime
with two independent angular momenta. However if one supplements an additional counterterm
into the usual Dirac operator, then the modified Dirac equation for the spin-$1/2$ spinor
particles is separable in this rotating, charged Einstein-Maxwell-Chern-Simons black hole
background geometry. A first-order symmetry operator that commutes with the modified Dirac
operator has exactly the same form as that previously found in the uncharged Myers-Perry
black hole case. It is expressed in terms of a rank-three totally antisymmetric tensor
and its covariant derivative. This tensor obeys a generalized Killing-Yano equation and its
square is a second-order symmetric St\"{a}ckel-Killing tensor admitted by the five-dimensional
rotating, charged black hole spacetime.
\end{abstract}

% 04.20.Jb, 03.65.Pm

\maketitle
%\end{CJK*}

%%%%%%%%%%%%%%%%%%%%%%%
\section{Introduction}
%%%%%%%%%%%%%%%%%%%%%%%

Soon after Chandrasekhar \cite{SC} showed that the massive Dirac's equation can be separated by
variables in the Kerr geometry \cite{RK} by using the null tetrad formalism, Page \cite{DonP}
and Lee \cite{CHL} demonstrated that the same thing holds true in the case of a Kerr-Newman black
hole \cite{KNm}. These works were extended in \cite{ISdyon} to the case of a rotating, charged
dyonic black hole background. Carter and McLenaghan \cite{CM} further found that the separability
of Dirac's equation in the Kerr geometry is related to the fact that the skew-symmetric tensor
admitted by the Kerr metric is a Killing-Yano tensor of rank-two, and its square is just a
second-order symmetric St\"{a}ckel-Killing tensor discovered by Carter \cite{BC}. Specifically speaking,
they established the correspondence between the separation constant appearing in the separable
solutions to the Dirac equation and the first-order differential operator that commutes with the
Dirac operator. This first-order symmetry operator can be constructed from the antisymmetric
Killing-Yano tensor, which implies an additional integral of motion that can be physically associated
with angular momentum \cite{CM}. Therefore, the separation of Dirac's equations can be understood
in terms of this first-order differential operator that characterizes the separation constant
appearing in the separated Dirac equations. The essential property that allows the construction
of such a symmetry operator is the existence of a Killing-Yano tensor in the Kerr spacetime.
Subsequently, the most general symmetry operator commuting with the Dirac operator was obviously
constructed \cite{MSKM} in a four-dimensional case. Later, it was shown \cite{CDC} that many of the
remarkable properties of the Kerr spacetime are consequences of the existence of the Killing-Yano
tensor, which means that all the symmetries responsible for the separability of various wave
equations are ``derivable'' from the Killing-Yano tensor.

It is necessary to extend these studies to the case of rotating black holes in more than four dimensions
since higher-dimensional generalizations (with or without a cosmological constant) of the Kerr black hole
and their properties have attracted considerable attention in recent years, in particular, in the context
of string theory, with the discovery of the anti-de Sitter/conformal-field-theory (AdS/CFT) correspondence,
and with the advent of brane-world theories. In our previous work \cite{WuP1}, we have investigated the
separability of a massive fermion field equation in the five-dimensional Myers-Perry \cite{MP} spacetime
with two unequal angular momenta and its relation to a rank-three Killing-Yano tensor. A first-order
symmetry operator commuting with the Dirac operator has been constructed by using the rank-three
Killing-Yano tensor whose square is just the rank-two symmetric St\"{a}ckel-Killing tensor. In addition,
we have obtained a second-order symmetry operator that commutes with the scalar Laplacian operator.

These results have further extended previous studies of hidden symmetry and separability properties of
rotating vacuum spacetimes, where the whole business of separation of variables in higher dimensions
was initiated by Frolov and his collaborators \cite{FS} (see \cite{VFDK} for a review and references
therein). In particular, they \cite{KF} first introduced the concept of a so-called principal conformal
Killing-Yano tensor and showed that starting with this tensor, a tower of Killing objects can be
generated. Inspired from their remarkable work \cite{KF}, our construction of the dual first-order
differential symmetry operator from the expected Killing-Yano tensors was successfully achieved,
although a general theory for it in any dimensions was already presented in \cite{TC}.

In another paper \cite{WuP2}, we have further studied the relation between the separability of Dirac's
equation and the Killing-Yano tensor in the general five-dimensional Kerr-AdS black holes \cite{HHT}
with two independent angular momenta. It has been shown that the separability of the massive Klein-Gordon
scalar field equation is intimately connected with the existence of a second-order St\"{a}ckel-Killing
tensor admitted by the five-dimensional Kerr-AdS spacetime. It is further demonstrated that this rank-two
St\"{a}ckel-Killing tensor can be constructed from its ``square root,'' a rank-three Killing-Yano tensor,
which is responsible for the separability of Dirac's equation. An easy way to obtain this tensor is that
one can start from a potential one-form to generate a rank-two conformal Killing-Yano tensor, whose Hodge
dual is just the expected Killing-Yano tensor.

The next step is natural to extend the above analysis to the case of charged generalizations of the
four-dimensional Kerr-Newman black hole in five dimensions. For this purpose, one should base upon
an analytical solution of the five-dimensional Kerr-Newman black hole. Unfortunately, an exact solution
of the rotating, charged Kerr-Newman black hole in higher dimensions still remains unknown up to now,
although the neutrally-charged generalizations of the Kerr metric to higher dimensions were obtained
\cite{MP} many years ago. So far, five-dimensional rotating charged black holes in pure Einstein-Maxwell
theory have only been studied numerically \cite{NS} and approximately \cite{AS}. It is likely that an
exact rotating charged solution does not exist at all in five dimensions within the pure Einstein-Maxwell
systems though static solutions in higher dimensions have been known for a long time. Evidence to
support this conjecture is that in the simplest case for $D = 5$ dimensions, the Einstein field
equation and the Maxwell equation can not simultaneously be satisfied \cite{AS} within such a pure
Einstein-Maxwell theory.

On the other hand, string theory provides powerful tools to generate fruitful solutions \cite{Youm}
for rotating charged black holes in higher dimensions. Motivated by ideas arising in various models
in string theory, supersymmetric rotating charged black holes were found \cite{BMPV,KS} within minimal
$D = 5$ supergravity theory. The existence of these solutions is made possible due to the peculiar
Chern-Simons coupling of minimal $D = 5$ supergravity and the fact that a self-duality condition is
allowed to be imposed on the exterior derivative of the rotation one-form in $D = 5$ dimensions. What
is more, the inclusion of a Chern-Simons term makes it easier to solve the field equations in the minimal
supergravity case \cite{KNL}. Without such an additional Chern-Simons term to the Maxwell equation,
it is very difficult to achieve an exact rotating charged solution to pure Einstein-Maxwell systems.

So far, all the exact solutions that have been found to describe rotating charged black holes in five
dimensions belong to the Einstein-Maxwell-Chern-Simons (EMCS) theory. Exact nonextremal solutions
representing five-dimensional rotating charged black holes with two equal angular momenta were also
found in AdS spaces \cite{CLP}. Rotating charged black holes embedded in the G\"{o}del universe have been
constructed quite recently \cite{GBH} in minimal $D = 5$ supergravity.

Recently, an exact solution describing a nonextremal rotating charged black hole in five-dimensional
AdS spaces was obtained \cite{CCLP} in minimally gauged supergravity. This solution is characterized
by the mass, the electric charge, the cosmological constant and two unequal angular momenta,
representing the most general charged generalizations of Kerr-AdS black holes in five dimensions.
By sending the cosmological constant to zero, one obtains a general nonextremal solution that
describes an asymptotically flat, rotating charged black hole, which can be viewed as an exact
charged generalization of the Kerr-Newman solution in five dimensions within the framework of
minimal $D = 5$ supergravity theory. This asymptotically flat metric is included as a special
case by setting three $U(1)$ charges equal in the general three-charge solutions that were generated
by Cveti\v{c} and Youm \cite{CY} in string theory more than ten years ago. The apparent relation
between them has been disclosed in Appendix B of Ref. \cite{BCCGS}. We shall call this
asymptotically flat solution for shortness the Cveti\v{c}-Youm or EMCS black hole and
consider it as the background metric in this paper. This general solution includes the famous
Breckenridge-Myers-Peet-Vafa (BMPV) \cite{BMPV} black hole as a special case when two rotation
parameters, the mass parameter and the electric charge simultaneously satisfy $b = \pm a$ and
$M = \mp Q$.

The main purpose of this article is to extend our previous research \cite{WuP1,WuP2} to deal with
the separation of a spin-$1/2$ spinor field in the five-dimensional rotating, charged Cveti\v{c}-Youm
black hole background spacetime \cite{CY} with two independent angular momenta and three equal
$U(1)$ charges. The black hole solution is a natural generalization of the famous Kerr-Newman solution
to five dimensions in $D = 5$ minimal supergravity theory. It is shown that the usual Dirac equation
can not be separated by variables in this general rotating, charged black hole spacetime with two
independent angular momenta. However, if one supplements an additional counterterm into the usual
Dirac operator, then the modified Dirac equation for the spin-$1/2$ spinor particles is separable
in this Cveti\v{c}-Youm black hole background geometry. A first-order symmetry operator commuting
with the modified Dirac operator has precisely the same form \cite{WuP1} as that previously found
in the uncharged Myers-Perry black hole case. It is expressed in terms of a rank-three totally
antisymmetric tensor and its covariant derivative. This tensor obeys a generalized Killing-Yano
equation and its square is a second-order symmetric St\"{a}ckel-Killing tensor admitted by this
general rotating, charged EMCS black hole spacetime. In addition, we have studied the separation
of variables for a massive Klein-Gordon equation in this five-dimensional Cveti\v{c}-Youm black
hole background spacetime. The symmetry operator that commutes with the scalar Laplacian operator is
directly constructed from the separated parts of the solutions and expressed in terms of the
rank-two St\"{a}ckel-Killing tensor.

Two pieces of previous work should be mentioned about the separation of the Dirac equation in
higher-dimensional rotating vacuum spacetimes and in a rotating charged black hole within minimal $D = 5$
supergravity. In Ref. \cite{OY}, the separability of the Dirac equation in arbitrary dimensional vacuum
black hole spacetimes was published prior to our work \cite{WuP1,WuP2}. However, the separability
of the Dirac equation and its connection with the Killing-Yano tensors had not obviously been demonstrated
there. What is more, the authors had not explicitly presented the symmetry operator that commutes
with the Dirac operator, using the Killing-Yano tensors. On the other hand, the authors of Ref.
\cite{DKL} had studied the separability of the Hamilton-Jacobi and Klein-Gordon equations in the $D = 5$
rotating charged Kerr-AdS black hole spacetime \cite{CCLP} of minimal gauged supergravity, and
presented the St\"{a}ckel-Killing tensors. However, they were only able to separate the usual Dirac
equation in the case of equal rotation parameters. They failed to find the ``generalized'' Killing-Yano
tenors (first given in this paper) and construct the dual symmetry operator. So, the whole hidden
symmetries for the $D = 5$ EMCS black hole in terms of generalized conformal Killing-Yano tenors had
not been completely revealed in their work. In the case of equal rotation parameters, their work
can naturally be recovered by ours if one omits the extra counterterm introduced in our paper.
However, even in this special case, the counterterm still makes an important contribution to the
theory. Therefore, their work on the separation of Dirac equation is incomplete in this sense.

The organization of this paper is outlined as follows. In Sec. \ref{CYBH}, a simple form for the
line element of the five-dimensional Cveti\v{c}-Youm black hole is presented in the Boyer-Lindquist
coordinates, with which we can explicitly construct the local orthonormal coframe one-forms (pentad)
just like the uncharged case \cite{WuP1}. In Sec. \ref{KGSKT}, we focus on the separation of
variables for a massive Klein-Gordon equation in this background and use the separated solutions
to construct a concise expression for the St\"{a}ckel-Killing tensor and a second-order operator
that commutes with the scalar Laplacian operator. Section \ref{mDirac} is devoted to the separation
of variables for a modified Dirac's equation in this five-dimensional EMCS black hole geometry. In
this section, we show that the usual massive Dirac equation can not be separated into purely radial
and purely angular parts, unless one supplements an additional counterterm into the usual Dirac
equation. With the inclusion of this new counterterm, the modified Dirac equation for a spin-$1/2$
spinor field in this general rotating, charged EMCS black hole spacetime can be completely decoupled
into purely radial and purely angular parts. In Sec. \ref{dualDirac}, we shall demonstrate that the
separated parts of the modified Dirac's equation also allow us to construct a first-order symmetry
operator that commutes with the modified Dirac operator. This operator has exactly the same form
\cite{WuP1} as that previously found in the uncharged Myers-Perry black hole case and is explicitly
expressed in terms of a rank-three totally antisymmetric tensor (and its covariant derivative)
admitted by the five-dimensional EMCS black hole spacetime. This fact implies that the inclusion
of a new additional counterterm is reasonable and our modification is compatible with our previous
results \cite{WuP1,WuP2}. It is easy to check that this antisymmetric tensor of rank-three does not
satisfy the usual Killing-Yano equation but a generalized form for it, while the Hodge dual of this
tensor can be generated from a potential one-form that has an expression similar to the uncharged
case \cite{WuP1,WuP2}. Section \ref{CoRe} ends with a brief summary of this paper and the related
work. In the Appendix, the affine spin-connection one-forms are calculated by the first Cartan
structure equation from the exterior differential of the pentad. The corresponding spinor-connection
one-forms, the curvature two-forms, and various useful tensors are also given in this pentad
formalism.

%%%%%%%%%%%%%%%%%%%%%%%%%%%%%%%%%%%%%%%%%%%%
\section{The Cveti\v{c}-Youm black hole solution
with three equal $U(1)$ charges} \label{CYBH}
%%%%%%%%%%%%%%%%%%%%%%%%%%%%%%%%%%%%%%%%%%%%

The five-dimensional rotating, charged black hole metric considered in this paper is a special case
of the general three-charge solutions constructed by Cveti\v{c} and Youm \cite{CY} in string theory,
within which three $U(1)$ charges are set equal. After doing that, the solution is greatly simplified
and obeys the complete Einstein-Maxwell-Chern-Simons equations in $D = 5$ minimal (ungauged)
supergravity theory. The bosonic part of this theory consists of the metric and a one-form gauge field,
which are governed by the EMCS equations of motion derived from the action
\be
S = \frac{1}{16\pi}\int d^5x \Big[\sqrt{-g}\big(R -F_{\mu\nu}F^{\mu\nu}\big)
 -\frac{2}{3\sqrt{3}}\epsilon^{\mu\nu\alpha\beta\gamma}F_{\mu\nu}
 F_{\alpha\beta}\mathcal{A}_{\gamma}\Big] \, .
\ee
The Einstein equation and the Maxwell-Chern-Simons equation read
\bea
&& R_{\mu\nu} -\frac{1}{2}g_{\mu\nu}R = 2T_{\mu\nu}
  \equiv 2\Big(F_{\mu\alpha}F_{\nu}^{~\alpha}
  -\frac{1}{4}g_{\mu\nu}F_{\alpha\beta}F^{\alpha\beta}\Big) \, , \qquad \\
&& \p_{\nu}\Big(\sqrt{-g}F^{\mu\nu} +\frac{1}{\sqrt{3}}
  \epsilon^{\mu\nu\alpha\beta\gamma} \mathcal{A}_{\alpha}F_{\beta\gamma}\Big) = 0 \, .
\eea
One can include a negative cosmological constant into this theory so that it becomes the minimal
$D = 5$ gauged supergravity theory. A general solution that describes a $D = 5$ non-extremal rotating,
charged black hole with two independent angular momenta and a negative cosmological constant has been
given in Ref. \cite{CCLP} and various aspects of rotating charged EMCS black holes have been studied
recently in \cite{DKL,EMCS,TYI}. Sending the cosmological constant to zero, the metric recovers the
Cveti\v{c}-Youm black hole solution in which three $U(1)$ charges are set equal. The correspondence
between these solutions has been confirmed in the second appendix of Ref. \cite{BCCGS}.

The metric and the gauge potential presented below simultaneously solve the Einstein equation and the
Maxwell-Chern-Simons equation. For our purpose in this paper, we find that the metric for the $D = 5$
EMCS black hole can be diagonalized into a very convenient form in order to construct a local orthonormal
pentad with which the spinor field equation can be decoupled into purely radial and purely angular
parts. As in Ref. \cite{WuP1}, we find that this line element can be recast into a simple form in
terms of the Boyer-Lindquist coordinates as follows:
\bea
ds^2 &=& g_{\mu\nu}dx^{\mu}dx^{\nu} = \eta_{AB}e^A\otimes e^B \,  \nn \\
 &=& -\frac{\Delta_r}{\Sigma} X^2 +\frac{\Sigma}{\Delta_r}dr^2 +\Sigma d\theta^2
  +\frac{(a^2-b^2)^2\sin^2\theta\cos^2\theta}{p^2\Sigma} Y^2
  +\Big(\frac{ab}{rp} Z +\frac{Qp}{r\Sigma} X\Big)^2 \, ,
\label{CY1c}
\eea
and the gauge potential is
\be
\mathcal{A} = \frac{\sqrt{3}Q}{2\Sigma} X \, ,
\ee
where
\bea
X &=& dt -a\sin^2\theta d\phi -b\cos^2\theta d\psi \, , \nn \\
Y &=& dt -\frac{(r^2+a^2)a}{a^2-b^2}d\phi -\frac{(r^2+b^2)b}{b^2-a^2}d\psi \, , \\
Z &=& dt -\frac{r^2+a^2}{a}\sin^2\theta d\phi -\frac{r^2+b^2}{b}\cos^2\theta d\psi \, , \nn
\eea
and
\be
r^2\Delta_r = (r^2+a^2)(r^2+b^2) -2Mr^2 +Q^2+2Qab \, , \quad
\Sigma = r^2 +p^2 \, , \quad~~ p = \sqrt{a^2\cos^2\theta +b^2\sin^2\theta} \, .
\ee
Here the parameters ($M, Q, a, b$) are related to the mass, the electric charge, and two independent
angular momenta of the black hole. Our conventions are as follows: Greek letters $\mu, \nu$ run over
five-dimensional spacetime coordinate indices $\{t, r, \theta, \phi, \psi\}$, while Latin letters $A,
B$ denote local orthonormal (Lorentzian) frame indices $\{0, 1, 2, 3, 5\}$. $\eta_{AB} = diag (-1, 1, 1,
1, 1)$ is the flat (Lorentzian) metric tensor. Units are used as $G = \hbar = c = 1$ from the beginning.

An important case with special interest is the supersymmetric BMPV black hole solution \cite{BMPV}.
It is included as a special case when the black hole becomes extremal, and the angular velocities at
the horizon vanishes. Equivalently, we have $b = \epsilon a$, and $M = -\epsilon Q$, where $\epsilon
= \pm 1$. To see this, we now let $b = \epsilon a$ and $\rho^2 = r^2 +a^2$. After some simplifications
we get
\bea
ds^2 &=& -\frac{V}{\rho^2-a^2}\big(dt -a\sigma_3\big)^2 +\frac{\rho^2}{V}d\rho^2
 +\rho^2d\theta^2 +\rho^2\sin^2\theta\cos^2\theta\big(d\phi -\epsilon d\psi\big)^2 \nn \\
 && +\frac{1}{\rho^2-a^2}\Big[adt -\rho^2\sigma_3 +\frac{\epsilon Qa}{\rho^2}\big(dt
 -a\sigma_3\big)\Big]^2 \, , \label{Bmab} \\
\mathcal{A} &=& \frac{\sqrt{3}Q}{2\rho^2} \big(dt -a\sigma_3\big) \, ,
\eea
where
\be
V = \rho^2 -2M +\frac{Q^2 +2(M +\epsilon Q)a^2}{\rho^2} \, , \qquad
\sigma_3 = \sin^2\theta d\phi +\epsilon\cos^2\theta d\psi \, .
\ee
Further adopting the Euler angle coordinates, the above line element and the gauge potential
can be recast into the general ansatz for the metric and potential given in \cite{GBH}. The
explicit solution is given there by sending the G\"{o}del parameter to zero.

It is known that a supersymmetric black hole must be an extremal one, but the converse is
not true in general \cite{EEF}. The extremal limit is $M^2 -Q^2 = 2(M +\epsilon Q)a^2$. When
the condition $M +\epsilon Q = 0$ holds, the angular velocities at the degenerate horizon
$\rho_e^2 = M$ vanish. It is easy to show that the only extremal solution with regular
horizon and vanishing angular velocities takes place under the conditions $b = \epsilon a$
and $M = -\epsilon Q$, namely, the BMPV solution. On the other hand, it has already been
proven in \cite{HSR} that the BMPV solution is the only asymptotically flat, supersymmetric,
rotating charged black hole with a regular, finite size horizon and finite entropy. It
is reasonable to infer that the Cveti\v{c}-Youm with three equal $U(1)$ charges must also
be unique. Recently, the uniqueness aspect of rotating charged black holes has been
discussed \cite{TYI,AA} in five-dimensional minimal gauged supergravity. This fact implies
that we might have little hope in finding an exact rotating charged black hole solution of
the pure Einstein-Maxwell system in five dimensions, although many people think such a kind
of solution should exist in higher dimensions and may be very complicated to construct
analytically.

Our interest in this paper is the general non-extremal case with two unequal rotation parameters.
The outer event horizon is determined by the largest root of $\Delta_{r_+} = 0$. The Hawking
temperature $T = \kappa/(2\pi)$ and the Bekenstein-Hawking entropy $S = A/4$ with respect to
this horizon can be easily computed as
\be
T = \frac{r_+^4 -(Q +ab)^2}{2\pi
 r_+\big[(r_+^2 +a^2)(r_+^2 +b^2) +Qab\big]} \, , \qquad
S = \pi^2\frac{(r_+^2 +a^2)(r_+^2+b^2) +Qab}{2r_+} \, ,
\ee
while the angular velocities and the electrostatic potential are measured relative to the observer
at infinity as
\bea
&& \Omega_a = \frac{a(r_+^2 +b^2) +Qb}{(r_+^2 +a^2)(r_+^2 +b^2) +Qab} \, , \qquad
\Omega_b = \frac{b(r_+^2 +a^2) +Qa}{(r_+^2 +a^2)(r_+^2 +b^2) +Qab} \, , \nn \\
&&~ \Phi = \frac{\sqrt{3}Qr_+^2/2}{(r_+^2 +a^2)(r_+^2+b^2) +Qab} \, .
\eea

The physical mass, two angular momenta, and the electric charge are given by
\be
\mathcal{M} = \frac{3\pi}{4}M \, , \qquad
J_a = \frac{\pi}{4}\big(2Ma +Qb\big) \, , \qquad
J_b = \frac{\pi}{4}\big(2Mb +Qa\big) \, , \qquad
\mathcal{Q} = \frac{\sqrt{3}\pi}{2}Q \, ,
\ee
which obey the closed forms for the first law of black hole thermodynamics
\bea
\frac{2}{3}\mathcal{M} &=& TS +\Omega_aJ_a +\Omega_bJ_b +\frac{2}{3}\Phi\mathcal{Q} \, , \\
d\mathcal{M} &=& TdS +\Omega_adJ_a +\Omega_bdJ_b +\Phi d\mathcal{Q} \, .
\eea

In practice, it is much more efficient to use $p$ rather than $\theta$ itself as the appropriate
angle coordinate. What is more, the radial part and the angular part can be presented in a symmetric
manner. In what follows, we shall adopt $p$ as the convenient angle variable throughout this article.
In doing so, the five-dimensional Cveti\v{c}-Youm black hole metric can be rewritten as
\be
ds^2 = -\frac{\Delta_r}{\Sigma} X^2 +\frac{\Sigma}{\Delta_r}dr^2
  +\frac{\Sigma}{\Delta_p}dp^2 +\frac{\Delta_p}{\Sigma} Y^2
  +\Big(\frac{ab}{rp} Z +\frac{Qp}{r\Sigma} X\Big)^2 \, ,
\label{mPDf}
\ee
where
\be
\Delta_p = -(p^2-a^2)(p^2-b^2)/p^2 \, ,
\ee
and
\bea
X &=& dt -\frac{(p^2-a^2)a}{b^2-a^2}d\phi
 -\frac{(p^2-b^2)b}{a^2-b^2}d\psi \, , \nn \\
Y &=& dt +\frac{(r^2+a^2)a}{b^2-a^2}d\phi
 +\frac{(r^2+b^2)b}{a^2-b^2}d\psi\, , \\
Z &=& dt -\frac{(r^2+a^2)(p^2-a^2)}{(b^2-a^2)a}d\phi
 -\frac{(r^2+b^2)(p^2-b^2)}{(a^2-b^2)b}d\psi\, . \nn
\eea

Completing the following coordinate transformations:
\be
t = \tau +(a^2+b^2)u +a^2b^2v \, , \qquad \phi = a(u +b^2v) \, , \qquad
\psi = b(u +a^2v) \, ,
\ee
we get
\bea
X = d\tau +p^2du \, , \qquad Y = d\tau -r^2du \, , \qquad
Z = d\tau +(p^2-r^2)du -r^2p^2dv \, .
\eea
The line element (\ref{mPDf}) is also applicable to the general solution of $D = 5$ rotating
charged Kerr-AdS black holes found in \cite{CCLP}. The similar metric form was also adopted in
\cite{LMP} recently. Soon after the discovery of the solution constructed in \cite{CCLP}, the
above elegant expressions for these metrics had already been obtained by the present author
for the purpose of considering the separation of the massive Dirac equation in these background
spacetimes. However, a satisfactory answer to the problem was achieved a little more than one
year ago.

The Cveti\v{c}-Youm black hole metric (\ref{CY1c}) possesses three Killing vectors ($\p_t$, $\p_{\phi}$,
and $\p_{\psi}$). In addition, it also admits a rank-two symmetric St\"{a}ckel-Killing tensor, which
can be written as the square of a rank-three antisymmetric tensor that obeys a generalized Killing-Yano
equation. In this paper, we will show that the existence of the St\"{a}ckel-Killing tensor ensures
the separation of variables in a massive Klein-Gordon scalar field equation, and the separability
of a modified Dirac's equation in this spacetime background is also closely associated with the
existence of a generalized Killing-Yano tensor of rank-three.

In five dimensions, there exist two different schemes \cite{PJDS,TypeD} for the algebra classification
of Weyl curvature tensors. According to the Weyl spinor classification \cite{PJDS}, the Myers-Perry
and Kerr-AdS black holes \cite{MP,HHT} are of Petrov type $\underline{22}$, while the BMPV solution
is of Petrov type 22. In the Weyl tensor classification scheme \cite{TypeD}, the former uncharged
vacuum solutions are of type D, and the latter supersymmetric BMPV black hole is of type I$_i$. It
is anticipated that the general spacetime metric (\ref{CY1c}) is of Petrov type 22 (or I$_i$), like
its supersymmetric charged version. This spacetime possesses a pair of real repeated principal null
vectors $\{\mathbf{l}, \mathbf{n}\}$, a pair of complex repeated principal null vectors $\{\mathbf{m},
\bar{\mathbf{m}}\}$, and one real, spatial-like unit vector $\mathbf{k}$. They can be constructed to
be of Kinnersley-type as follows:
\bea
\mathbf{l}^{\mu}\p_{\mu} &=& \frac{1}{r^2\Delta_r}\Big[(r^2+a^2)(r^2+b^2)\p_t
 +(r^2+b^2)a\p_{\phi} +(r^2+a^2)b\p_{\psi} \nn \\
&&\quad +Q\big(ab\p_t +b\p_{\phi} +a\p_{\psi}\big)\Big] +\p_r \, , \nn \\
\mathbf{n}^{\mu}\p_{\mu} &=& \frac{1}{2r^2\Sigma}\Big[(r^2+a^2)(r^2+b^2)\p_t
 +(r^2+b^2)a\p_{\phi} +(r^2+a^2)b\p_{\psi} \nn \\
&&\quad +Q\big(ab\p_t +b\p_{\phi} +a\p_{\psi}\big)\Big]
 -\frac{\Delta_r}{2\Sigma}\p_r \, , \nn \\
\mathbf{m}^{\mu}\p_{\mu} &=& \frac{\sqrt{\Delta_p/2}}{r+ip}\bigg[\p_p
 +i\frac{(p^2-a^2)(p^2-b^2)}{p^2\Delta_p}\Big(\p_t
 -\frac{a}{p^2-a^2}\p_{\phi} -\frac{b}{p^2-b^2}\p_{\psi}\Big)\bigg] \, , \nn \\
\bar{\mathbf{m}}^{\mu}\p_{\mu} &=& \frac{\sqrt{\Delta_p/2}}{r-ip}\bigg[\p_p
 -i\frac{(p^2-a^2)(p^2-b^2)}{p^2\Delta_p}\Big(\p_t
 -\frac{a}{p^2-a^2}\p_{\phi} -\frac{b}{p^2-b^2}\p_{\psi}\Big)\bigg] \, , \nn \\
\mathbf{k}^{\mu}\p_{\mu} &=& \frac{1}{rp}\big(ab\p_t +b\p_{\phi} +a\p_{\psi}\big) \, .
\eea
These vectors satisfy the following orthogonal relations
\be
\mathbf{l}^{\mu}\mathbf{n}_{\mu} = -1 \, ,  \qquad\quad \mathbf{m}^{\mu}\bar{\mathbf{m}}_{\mu} = 1 \, ,
\qquad\quad \mathbf{k}^{\mu}\mathbf{k}_{\mu} = 1 \, ,
\label{Ognr}
\ee
and all others are zero. In terms of these vectors, the metric for the Cveti\v{c}-Youm black hole
(\ref{CY1c}) can be put into a seminull pentad formalism ($2\bar{2}1$ formalism \cite{WuP1,WuP2})
as follows:
\be
ds^2 = -\mathbf{l}\otimes \mathbf{n} -\mathbf{n}\otimes \mathbf{l} +\mathbf{m}\otimes
 \bar{\mathbf{m}} +\bar{\mathbf{m}}\otimes \mathbf{m} +\mathbf{k}\otimes \mathbf{k} \, .
\ee

%%%%%%%%%%%%%%%%%%%%%%%%%%%%%%%%%%%%%%%%%%%%%%%%%%%%%%%%%%%%%%%%
\section{St\"{a}ckel-Killing tensor and second-order symmetry
operator from the separated Klein-Gordon equation} \label{KGSKT}
%%%%%%%%%%%%%%%%%%%%%%%%%%%%%%%%%%%%%%%%%%%%%%%%%%%%%%%%%%%%%%%%

In this section, the massive Klein-Gordon scalar field equation is shown to be separable in the
five-dimensional Cveti\v{c}-Youm metric. From the separated solution of the radial and angular
parts, we can construct a second-order symmetry operator that commutes with the scalar Laplacian
operator. We show that a second-order, symmetric, St\"{a}ckel-Killing tensor has a concise
form in the local Lorentzian pentad, which can be easily written as the square of a rank-three
generalized Killing-Yano tensor given in Sec. \ref{dualDirac}.

To begin with, let us consider a massive Klein-Gordon scalar field equation
\be
\big(\Box -\mu_0^2\big)\Phi = \frac{1}{\sqrt{-g}}\p_{\mu}
 \big(\sqrt{-g} g^{\mu\nu}\p_{\nu}\Phi\big) -\mu_0^2\Phi = 0 \, ,
\ee
where $\mu_0$ is the rest mass of the scalar particle.

The metric determinant for this spacetime is $\sqrt{-g} = rp\Sigma/(a^2-b^2)$, and the
contrainvariant metric tensor can be read accordingly from
\bea
g^{\mu\nu}\p_{\mu}\p_{\nu} &=& \eta^{AB}\p_A\otimes\p_B \nn \\
&=& -\frac{1}{r^4\Delta_r\Sigma}\Big[(r^2+a^2)(r^2+b^2)\p_t
 +(r^2+b^2)a\p_{\phi} +(r^2+a^2)b\p_{\psi} \nn \\
&& +Q\big(ab\p_t +b\p_{\phi} +a\p_{\psi}\big)\Big]^2
 +\frac{\Delta_r}{\Sigma}\p_r^2 +\frac{\Delta_p}{\Sigma}\p_p^2 \nn \\
&& +\frac{(p^2-a^2)^2(p^2-b^2)^2}{p^4\Delta_p\Sigma}
\Big(\p_t -\frac{a}{p^2-a^2}\p_{\phi} -\frac{b}{p^2-b^2}\p_{\psi}\Big)^2 \nn \\
&&\quad +\frac{1}{r^2p^2}\big(ab\p_t +b\p_{\phi} +a\p_{\psi}\big)^2 \, .
\eea

The massive scalar field equation in the background spacetime metric (\ref{CY1c}) can be
explicitly written as
\bea
&&\bigg\{ -\frac{1}{r^4\Delta_r\Sigma}\Big[(r^2+a^2)(r^2+b^2)\p_t
 +(r^2+b^2)a\p_{\phi} +(r^2+a^2)b\p_{\psi} \nn \\
&&\qquad +Q\big(ab\p_t +b\p_{\phi} +a\p_{\psi}\big)\Big]^2
 +\frac{1}{r\Sigma}\p_r\big(r\Delta_r\p_r\big)
  +\frac{1}{p\Sigma}\p_p\big(p\Delta_p\p_p\big) \nn \\
&&\qquad +\frac{(p^2-a^2)^2(p^2-b^2)^2}{p^4\Delta_p\Sigma}\Big(\p_t
 -\frac{a}{p^2-a^2}\p_{\phi} -\frac{b}{p^2-b^2}\p_{\psi}\Big)^2 \nn \\
&&\qquad\quad +\frac{1}{r^2p^2}\big(ab\p_t +b\p_{\phi} +a\p_{\psi}\big)^2
 -\mu_0^2\bigg \}\Phi = 0 \, .
\eea
Adopting the ansatz of separation of variables $\Phi = R(r)S(p)e^{i(m\phi +k\psi -\omega t)}$,
we can separate it into a radial part and an angular part,
\bea
&& \frac{1}{r}\p_r\big(r\Delta_r\p_rR\big) +\Big\{\frac{1}{r^4\Delta_r}
 \Big[(r^2+a^2)(r^2+b^2)\omega -(r^2+b^2)ma -(r^2+a^2)kb \nn \\
&&\qquad\qquad +Q\big(ab\omega -mb -ka\big)\Big]^2 -\frac{1}{r^2}\big(ab\omega
 -mb -ka\big)^2 -\mu_0^2r^2 -\lambda_0^2 \Big\}R(r) = 0 \, , \label{srs} \\
&& \frac{1}{p}\p_p\big(p\Delta_p\p_pS\big)
 -\Big\{\frac{(p^2-a^2)^2(p^2-b^2)^2}{p^4\Delta_p}\Big(\omega
 +\frac{ma}{p^2-a^2} +\frac{kb}{p^2-b^2}\Big)^2 \nn \\
&&\qquad\qquad\qquad\quad  +\frac{1}{p^2}\big(ab\omega -mb -ka\big)^2
 +\mu_0^2p^2 -\lambda_0^2\Big\}S(p) = 0 \, . \label{sra}
\eea

Now from the separated parts (\ref{srs}) and (\ref{sra}), one can construct a new dual equation
as follows:
\bea
&&\bigg\{ -\frac{p^2}{r^4\Delta_r\Sigma}\Big[(r^2+a^2)(r^2+b^2)\p_t
 +(r^2+b^2)a\p_{\phi} +(r^2+a^2)b\p_{\psi} \nn \\
&&\quad +Q\big(ab\p_t +b\p_{\phi} +a\p_{\psi}\big)\Big]^2
 +\frac{p^2}{r\Sigma}\p_r\big(r\Delta_r\p_r\big)
 -\frac{r^2}{p\Sigma}\p_p\big(p\Delta_p\p_p\big) \nn \\
&&\qquad -r^2\frac{(p^2-a^2)^2(p^2-b^2)^2}{p^4\Delta_p\Sigma}\Big(\p_t
 -\frac{a}{p^2-a^2}\p_{\phi} -\frac{b}{p^2-b^2}\p_{\psi}\Big)^2 \nn \\
&&\qquad\quad +\frac{p^2-r^2}{r^2p^2}\big(ab\p_t +b\p_{\phi}
 +a\p_{\psi}\big)^2 -\lambda_0^2\bigg \}\Phi = 0 \, ,
\eea
from which we can extract a second-order symmetric tensor --- the so-called St\"{a}ckel-Killing tensor
\bea
K^{\mu\nu}\p_{\mu}\p_{\nu} &=& -p^2\frac{1}{r^4\Delta_r\Sigma} \Big[(r^2+a^2)(r^2+b^2)\p_t
 +(r^2+b^2)a\p_{\phi} +(r^2+a^2)b\p_{\psi} \nn \\
&& +Q\big(ab\p_t +b\p_{\phi} +a\p_{\psi}\big)\Big]^2
 +p^2\frac{\Delta_r}{\Sigma}\p_r^2 -r^2\frac{\Delta_p}{\Sigma}\p_p^2 \nn \\
&& -r^2\frac{(p^2-a^2)^2(p^2-b^2)^2}{p^4\Delta_p\Sigma}\Big(\p_t
 -\frac{a}{p^2-a^2}\p_{\phi} -\frac{b}{p^2-b^2}\p_{\psi}\Big)^2 \nn \\
&&\quad +\frac{p^2-r^2}{r^2p^2}\big(ab\p_t +b\p_{\phi} +a\p_{\psi}\big)^2 \, .
\label{5dKT}
\eea
This symmetric tensor $K_{\mu\nu} = K_{\nu\mu}$ obeys the Killing equation
\be
K_{\mu\nu;\rho} +K_{\nu\rho;\mu} +K_{\rho\mu;\nu} = 0 \, .
\label{Kte}
\ee
In the local Lorentzian coframe given below in Eq. (\ref{pentad}), it has a simple, diagonal form
\be
K_{AB} = \mbox{diag} (-p^2, p^2, -r^2, -r^2, p^2-r^2) \, ,
\label{SKdf}
\ee
similar to that found in the uncharged case \cite{WuP1}.

Using the St\"{a}ckel-Killing tensor, the above dual equation can be written in a coordinate-independent
form
\be
\big(\mathbb{K} -\lambda_0^2\big)\Phi = \frac{1}{\sqrt{-g}}\p_{\mu}
 \big(\sqrt{-g} K^{\mu\nu}\p_{\nu}\Phi\big) -\lambda_0^2\Phi = 0 \, .
\ee
Clearly, the symmetry operator $\mathbb{K}$ is expressed in terms of the St\"{a}ckel-Killing tensor and
commutes with the scalar Laplacian operator $\Box$. Expanding the commutator $[\mathbb{K}, \Box] = 0$
yields the Killing Eq. (\ref{Kte}) and the integrability condition for the St\"{a}ckel-Killing
tensor.

%%%%%%%%%%%%%%%%%%%%%%%%%%%%%%%%%%%%%%%%%%%%%%%%%%%%%%%%%%
\section{Separability of the modified Dirac field equation
in a five-dimensional Cveti\v{c}-Youm black hole} \label{mDirac}
%%%%%%%%%%%%%%%%%%%%%%%%%%%%%%%%%%%%%%%%%%%%%%%%%%%%%%%%%%

In Ref. \cite{WuP1}, the usual Dirac equation for spin-$1/2$ fermions in the general Myers-Perry black
hole geometry has been decoupled into purely radial and purely angular parts by using the orthonormal
f\"{u}nfbein (pentad) formalism. In this section, we shall extend that work to the case of a rotating,
charged Cveti\v{c}-Youm black hole. To do this, we will cope with the spin-$1/2$ spinor field equation
within a local orthonormal pentad formalism and show that the modified Dirac equation is separable by
variables in the $D = 5$ Cveti\v{c}-Youm black hole geometry.

In five-dimensional curved background spacetime, the Dirac equation for the spin-$1/2$ spinor field is
\be
\big(\mathbb{H}_D +\mu_e\big)\Psi =
 \big[\gamma^Ae_A^{~\mu}(\p_{\mu} +\Gamma_{\mu}) +\mu_e\big]\Psi = 0 \, ,
\label{DE}
\ee
where $\Psi$ is a four-component Dirac spinor, $\mu_e$ is the rest mass of the electron, $e_A^{~\mu}$
is the f\"{u}nfbein (pentad), its inverse $e_{~\mu}^A$ is defined by $g_{\mu\nu} = \eta_{AB}e_{~\mu}^A
e_{~\nu}^B$, $\Gamma_{\mu}$ is the spinor connection, and $\gamma^A$'s are the five-dimensional gamma
matrices obeying the anticommutation relations (Clifford algebra)
\be
\big\{\gamma^A, \gamma^B\big\} \equiv \gamma^A\gamma^B +\gamma^B\gamma^A = 2\eta^{AB} \, .
\label{Clifford}
\ee
It is convenient to choose the following explicit representations for the gamma matrices
\bea
&& \gamma^0 = i\sigma^1\otimes \Id \, , \qquad\quad \gamma^1 = -\sigma^2\otimes \sigma^3 \, ,
\qquad\quad \gamma^2 = -\sigma^2\otimes \sigma^1 \, , \nn \\
&& \gamma^3 = -\sigma^2\otimes \sigma^2 \, , \qquad\qquad\quad
\gamma^5 = \sigma^3\otimes \Id = -i\gamma^0\gamma^1\gamma^2\gamma^3 \, ,
\label{GMr}
\eea
where $\sigma^i$'s are the Pauli matrices, and $\Id$ is a $2 \times 2$ identity matrix, respectively.

In the f\"{u}nfbein formalism, the Dirac field Eq. (\ref{DE}) can be rewritten in the local
Lorentzian frame as \cite{WuP1,WuP2}
\be
\big(\mathbb{H}_D +\mu_e\big)\Psi = \big[\gamma^A(\p_A +\Gamma_A) +\mu_e\big]\Psi = 0 \, ,
\ee
where $\p_A = e_A^{~\mu}\p_{\mu}$ is the local partial differential operator and $\Gamma_A = e_A^{~\mu}
\Gamma_{\mu}$ is the component of the spinor connection projected in the local Lorentzian frame. Therefore,
in order to get the explicit expression of Dirac's equation, one needs to find $\p_A$ and $\Gamma_A$
first. Once the pentad coframe one-forms $e^A = e^A_{~\mu}dx^{\mu}$ have been concretely chosen, the
local differential operator $\p_A = e_A^{~\mu}\p_{\mu}$ can be determined via the orthogonal relations:
$e_A^{~\mu}e^B_{~\mu} = \delta_A^B$ and $e_A^{~\mu}e^A_{~\nu} = \delta^{\mu}_{\nu}$.

The new form of the five-dimensional Cveti\v{c}-Youm metric (\ref{CY1c}) admits the following local
Lorentzian basis one-forms (pentad) $e^{A}$ orthonormal with respect to $\eta_{AB}$,
\be
 e^0 = \sqrt{\frac{\Delta_r}{\Sigma}} X \, , \quad
 e^1 = \sqrt{\frac{\Sigma}{\Delta_r}}dr \, , \quad
 e^2 = \sqrt{\frac{\Sigma}{\Delta_p}}dp \, , \quad
 e^3 = \sqrt{\frac{\Delta_p}{\Sigma}} Y \, , \quad
 e^5 = -\Big(\frac{ab}{rp} Z +\frac{Qp}{r\Sigma} X\Big) \, ,
\label{pentad}
\ee
from which we can easily get the dual orthonormal basis one-vectors $\p_A $ as follows:

\bea
&& \p_0 = \frac{1}{r^2\sqrt{\Delta_r\Sigma}}\Big[(r^2+a^2)(r^2+b^2)\p_t
 +(r^2+b^2)a\p_{\phi} \nn \\
&&\qquad +(r^2+a^2)b\p_{\psi} +Q\big(ab\p_t +b\p_{\phi}
 +a\p_{\psi}\big)\Big] \, , \nn \\
&& \p_1 = \sqrt{\frac{\Delta_r}{\Sigma}}\p_r \, , \qquad\qquad
 \p_2 = \sqrt{\frac{\Delta_p}{\Sigma}}\p_p \, , \nn \\
&& \p_3 = \frac{(p^2-a^2)(p^2-b^2)}{p^2\sqrt{\Delta_p\Sigma}}\Big(\p_t
 -\frac{a}{p^2-a^2}\p_{\phi} -\frac{b}{p^2-b^2}\p_{\psi}\Big) \, , \nn \\
&& \p_5 = \frac{1}{rp}\big(ab\p_t +b\p_{\phi} +a\p_{\psi}\big) \, .
\eea
Therefore, the spinor differential operator is
\bea
&& \gamma^A\p_A = \gamma^0 \frac{1}{r^2\sqrt{\Delta_r\Sigma}}\Big[(r^2+a^2)(r^2+b^2)\p_t
 +(r^2+b^2)a\p_{\phi} +(r^2+a^2)b\p_{\psi} \nn \\
&&\qquad\qquad +Q\big(ab\p_t +b\p_{\phi} +a\p_{\psi}\big)\Big]
 +\gamma^1\sqrt{\frac{\Delta_r}{\Sigma}}\p_r
 +\gamma^2\sqrt{\frac{\Delta_p}{\Sigma}}\p_p \nn \\
&&\qquad\qquad +\gamma^3 \frac{(p^2-a^2)(p^2-b^2)}{p^2\sqrt{\Delta_p\Sigma}}\Big(\p_t
 -\frac{a}{p^2-a^2}\p_{\phi} -\frac{b}{p^2-b^2}\p_{\psi}\Big) \nn \\
&&\qquad\qquad +\gamma^5\frac{1}{rp}\big(ab\p_t +b\p_{\phi} +a\p_{\psi}\big) \, .
\label{spdo}
\eea

The next step is to compute the component $\Gamma_A$ of the spinor connection. In order to derive
the spinor connection one-forms $\Gamma = \Gamma_{\mu}dx^{\mu}\equiv \Gamma_Ae^A$, one can first
compute the spin-connection one-forms $\omega_{AB} = \omega_{AB\mu}dx^{\mu}\equiv \Upsilon_{ABC}e^C$ in
the orthonormal pentad coframe, which can be determined from Cartan's first structure equation
and the skew-symmetric condition
\be
de^A +\omega^A_{~B}\wedge e^B = 0 \, , \qquad\qquad
\omega_{AB} = \eta_{AC}\omega_{~B}^C = -\omega_{BA} \, .
\label{CFE}
\ee
One can utilize $\Gamma_{\mu} = (1/8)[\gamma^A,\gamma^B]\omega_{AB\mu} = (1/4)\gamma^A\gamma^B
\omega_{AB\mu}$ to construct $\Gamma$ from $\omega_{AB}$,
\be
\Gamma = \frac{1}{8}\big[\gamma^A, \gamma^B\big]\omega_{AB}
 = \frac{1}{4}\gamma^A\gamma^B\omega_{AB} = \frac{1}{4}\gamma^A\gamma^B\Upsilon_{ABC}e^C \, ,
\ee
from which $\Gamma_A = (1/4)\gamma^B\gamma^C\Upsilon_{BCA}$ can be read out. The explicit expressions
for $\omega^A_{~B}$ and $\Gamma_A$ are displayed in the Appendix.

For our final purpose, we have to get the following expression for $\gamma^A\Gamma_A$ subject to the
Cveti\v{c}-Youm metric (\ref{CY1c}),
\bea
\gamma^A\Gamma_A &=& \frac{1}{4}\gamma^A\gamma^B\gamma^C\Upsilon_{BCA} \nn \\
&=& \gamma^1\sqrt{\frac{\Delta_r}{\Sigma}}\Big(\frac{\Delta_r^{\prime}}{4\Delta_r}
 +\frac{1}{2r} +\frac{r -ip\gamma^5}{2\Sigma}\Big) +\gamma^2\sqrt{\frac{\Delta_p}{\Sigma}}
 \Big(\frac{\Delta_p^{\prime}}{4\Delta_p} +\frac{1}{2p}
 +\frac{p +ir\gamma^5}{2\Sigma}\Big) \nn \\
&& +\Big(\frac{Q+ab}{2r^2\Sigma} +\frac{ab}{2p^2\Sigma}\Big)i\gamma^0\gamma^1
 \big(r +ip\gamma^5\big) -\frac{Q}{2\Sigma^2}\big(ir\gamma^0\gamma^1
  +p\gamma^0\gamma^1\gamma^5\big) \, ,
\label{spga}
\eea
where a prime denotes the partial differential with respect to the coordinates $r$ or $p$.

One main result of this paper is to point out that the existence of last term in the expression of
$\gamma^A\Gamma_A$ spoils the separability of the usual Dirac equation. However, this unexpected term
can be cancelled by including a new counterterm
\be
\frac{1}{12\sqrt{3}}\gamma^A\gamma^B\gamma^C\widetilde{F}_{ABC} =
\frac{Q}{2\Sigma^2}\big(p\gamma^0\gamma^1 +r\gamma^2\gamma^3\big)\gamma^5
\equiv \frac{i}{4\sqrt{3}}\gamma^A\gamma^B F_{AB} \, .
\label{wuct}
\ee
With this additional counterterm supplemented into the ordinary covariant spinor differential operator
$\mathbb{H}_D = \gamma^A(\p_A +\Gamma_A)$, the modified Dirac equation for the spin-$1/2$ spinor
field can be satisfactorily decoupled in the five-dimensional rotating charged background spacetime
considered in this paper. What is more, as we will see in the next section, the dual operator commuting
with this modified Dirac operator has precisely the same form as the previous one \cite{WuP1} found in
the uncharged Myers-Perry black hole case. That means, the minor measure to remedy the separability of
the usual Dirac equation in this background metric is to supplement an additional counterterm into it.
Thus, it is reasonable to believe that our treatment is on the right course. Following the present work,
the authors of Ref. \cite{KKY} recently proposed to interpret the Hodge dual three-form of the Maxwell
field strength two-form as a torsion tensor. As a result of this ``formal'' geometric identification,
the modification of the Dirac operator is quite natural from the viewpoint of this generalized ``torsion.''

At this stage, a few comments are in order. Frankly speaking, our initial aim of adding an extra
counterterm to the usual Dirac equation is to find out how we can achieve the separability of variables
in the spinor field equation. However, this modification is not ``\emph{freely}'' made. It is based
upon the following considerations: (1) Gauge invariance and general covariance should be kept in the
modified spinor field equation; (2) The modified equation should naturally reduced to the uncharged
case; (3) The dual first-order differential operator should have the same form as that in the uncharged
case. In other words, the pentad components of the Killing-Yano tensor should have the same expressions
as those in the uncharged case. The previous definition of the Killing-Yano tensor should not be changed or
only a minor modification is needed; (4) Inclusion of the additional counterterm should be explained
on the basis of supersymmetry. However, to the best of our knowledge, this issue has not been addressed
in the existing literature. Besides, as mentioned above, the inclusion of this additional counterterm
can be geometrically understood \cite{KKY} as a natural consequence if one identifies the dual Maxwell
three-form with a generalized ``torsion'' tensor.

Another important issue needed to be pointed out is that the separability of the Dirac equation in
a stationary spacetime is coordinate (tetrad)-dependent. Whether or not the Dirac equation is separable
is determined at least by (i) choosing an appropriate tetrad; and (ii) taking a suitable representation
for the gamma matrices. According to our previous experience, the choice of a suitable tetrad is very
crucial for this purpose. In this paper, the success of the separation by variables for the modified
Dirac equation relies heavily on the fact that we have found a most convenient, orthonormal pentad system
(\ref{pentad}), since the gamma matrices are easily chosen. Other pentad systems can be considered
by taking the Lorentzian (f\"{u}nfbein) transformations of this orthonormal pentad system, but in the
meanwhile one has to make the corresponding transformations on the gamma matrices and Dirac spinor
components. Apart from these, it is not easy and impossible to find some different coordinates or choose
a different pentad that one can use to completely separate the usual Dirac equation. In the next section,
we will demonstrate that it is the modified Dirac operator not the usual one that can commute with the
symmetry operator $\mathbb{H}_f$. This indicates that the modified Dirac equation is separable, while
the usual Dirac equation is not, in the general case with two unequal rotation parameters.

Combining Eqs. (\ref{spdo}) and (\ref{spga}) with the counterterm (\ref{wuct}), we find that the modified
Dirac's covariant differential operator in the local Lorentzian frame is
\bea
\widetilde{\mathbb{H}}_D &=& \gamma^A(\p_A +\Gamma_A)
 +\frac{1}{12\sqrt{3}}\gamma^A\gamma^B\gamma^C\widetilde{F}_{ABC} \nn \\
&=& \gamma^0\frac{1}{r^2\sqrt{\Delta_r\Sigma}}\Big[(r^2+a^2)(r^2+b^2)\p_t
 +(r^2+b^2)a\p_{\phi} +(r^2+a^2)b\p_{\psi} \nn \\
&& +Q\big(ab\p_t +b\p_{\phi} +a\p_{\psi}\big)\Big]
 +\gamma^1\sqrt{\frac{\Delta_r}{\Sigma}}\Big(\p_r +\frac{\Delta_r^{\prime}}{4\Delta_r}
 +\frac{1}{2r} +\frac{r -ip\gamma^5}{2\Sigma}\Big) \nn \\
&& +\gamma^2\sqrt{\frac{\Delta_p}{\Sigma}}\Big(\p_p +\frac{\Delta_p^{\prime}}{4\Delta_p}
 +\frac{1}{2p} +\frac{p +ir\gamma^5}{2\Sigma}\Big)
 +\gamma^3\frac{(p^2-a^2)(p^2-b^2)}{p^2\sqrt{\Delta_p\Sigma}}\Big(\p_t
 -\frac{a}{p^2-a^2}\p_{\phi} \nn \\
&& -\frac{b}{p^2-b^2}\p_{\psi}\Big) +\gamma^5\frac{1}{rp}\big(ab\p_t
 +b\p_{\phi} +a\p_{\psi}\big) +\Big(\frac{Q+ab}{2r^2\Sigma}
 +\frac{ab}{2p^2\Sigma}\Big)i\gamma^0\gamma^1\big(r +ip\gamma^5\big) \, . \qquad
\label{mdop}
\eea

With the above preparation in hand, we are now in a position to decouple the modified Dirac equation
\be
\big(\widetilde{\mathbb{H}}_D +\mu_e\big)\Psi = \Big[\gamma^A(\p_A +\Gamma_A)
 +\frac{1}{12\sqrt{3}}\gamma^A\gamma^B\gamma^C\widetilde{F}_{ABC} +\mu_e\Big]\Psi = 0 \, .
\label{DELF}
\ee
Substituting Eq. (\ref{mdop}) into Eq. (\ref{DELF}), the modified Dirac equation in the five-dimensional
Cveti\v{c}-Youm metric reads

\bea
&& \bigg\{ \gamma^0\frac{1}{r^2\sqrt{\Delta_r\Sigma}}\Big[(r^2+a^2)(r^2+b^2)\p_t
+(r^2+b^2)a\p_{\phi} +(r^2+a^2)b\p_{\psi} +Q\big(ab\p_t +b\p_{\phi} +a\p_{\psi}\big)\Big] \nn \\
&&\qquad +\gamma^1\sqrt{\frac{\Delta_r}{\Sigma}}\Big(\p_r +\frac{\Delta_r^{\prime}}{4\Delta_r}
 +\frac{1}{2r} +\frac{r -ip \gamma^5}{2\Sigma}\Big) +\gamma^2\sqrt{\frac{\Delta_p}{\Sigma}}
 \Big[\p_p +\frac{\Delta_p^{\prime}}{4\Delta_p} \nn \\
&&\qquad +\frac{1}{2p} +\frac{i\gamma^5}{2\Sigma}\big(r -ip\gamma^5\big)\Big]
  +\gamma^3\frac{(p^2-a^2)(p^2-b^2)}{p^2\sqrt{\Delta_p\Sigma}}
 \Big(\p_t -\frac{a}{p^2-a^2}\p_{\phi} -\frac{b}{p^2-b^2}\p_{\psi}\Big) \nn \\
&&\qquad +\gamma^5\frac{1}{rp}\big(ab\p_t +b\p_{\phi} +a\p_{\psi}\big)
 +\Big(\frac{Q+ab}{2r^2\Sigma} +\frac{ab}{2p^2\Sigma}\Big)i\gamma^0\gamma^1
 \big(r +ip\gamma^5\big) +\mu_e \bigg\}\Psi = 0 \, .
\eea
Multiplying $(r -ip\gamma^5)\sqrt{r +ip\gamma^5} = \sqrt{\Sigma(r -ip\gamma^5)}$ by the left
to the above equation, and after some lengthy algebra manipulations we arrive at
\bea
&& \bigg\{ \gamma^0\frac{1}{r^2\sqrt{\Delta_r}}\Big[(r^2+a^2)(r^2+b^2)\p_t
+(r^2+b^2)a\p_{\phi} +(r^2+a^2)b\p_{\psi} +Q\big(ab\p_t +b\p_{\phi} +a\p_{\psi}\big)\Big] \nn \\
&&\qquad  +\gamma^1\sqrt{\Delta_r}\Big(\p_r
 +\frac{\Delta_r^{\prime}}{4\Delta_r} +\frac{1}{2r}\Big) +\gamma^2\sqrt{\Delta_p}\Big(\p_p
 +\frac{\Delta_p^{\prime}}{4\Delta_p} +\frac{1}{2p}\Big) \nn \\
&&\qquad +\gamma^3\frac{(p^2-a^2)(p^2-b^2)}{p^2\sqrt{\Delta_p}}\Big(\p_t
 -\frac{a}{p^2-a^2}\p_{\phi} -\frac{b}{p^2-b^2}\p_{\psi}\Big) +\Big(\frac{\gamma^5}{p}
 -\frac{i}{r}\Big)\big(ab\p_t +b\p_{\phi} \nn \\
&&\qquad +a\p_{\psi}\big) +\Big(\frac{Q+ab}{2r^2} +\frac{ab}{2p^2}\Big)i\gamma^0\gamma^1
 +\mu_e\big(r -ip\gamma^5\big) \bigg\}\big(\sqrt{r +ip\gamma^5}\Psi\big) = 0 \, .
\label{presde}
\eea

Now applying the explicit representation (\ref{GMr}) for the gamma matrices and adopting the following
ansatz \cite{WuP1,WuP2} for the separation of variables
\be
\sqrt{r +ip\gamma^5}\Psi = e^{i(m\phi+k\psi-\omega t)}\left(\ba{cl}
&\hspace*{-5pt} R_2(r)S_1(p) \\
&\hspace*{-5pt} R_1(r)S_2(p) \\
&\hspace*{-5pt} R_1(r)S_1(p) \\
&\hspace*{-5pt} R_2(r)S_2(p)
\ea\right) \, ,
\ee
we find that the modified Dirac equation in the five-dimensional Cveti\v{c}-Youm metric can be decoupled
into the purely radial parts and the purely angular parts
\bea
&& \sqrt{\Delta_r}\cD_r^-R_1 = \Big[\lambda +i\mu_er -\frac{Q+ab}{2r^2}
 -\frac{i}{r}\big(ab\omega -mb -ka\Big)\big]R_2 \, , \label{sdea} \\
&& \sqrt{\Delta_r}\cD_r^+R_2 = \Big[\lambda -i\mu_er -\frac{Q+ab}{2r^2}
+\frac{i}{r}\big(ab\omega -mb -ka\big)\Big]R_1 \, , \label{sdeb} \\
&& \sqrt{\Delta_p}\cL_p^+S_1 = \Big[\quad\lambda +\mu_ep +\frac{ab}{2p^2}
 +\frac{1}{p}\big(ab\omega -mb -ka\big)\Big]S_2 \, , \label{sdec} \\
&& \sqrt{\Delta_p}\cL_p^-S_2 = \Big[-\lambda +\mu_ep -\frac{ab}{2p^2}
 +\frac{1}{p}\big(ab\omega -mb -ka\big)\Big]S_1 \label{sded} \, ,
\eea
in which
\bea
&& \cD_r^{\pm} = \p_r +\frac{\Delta_r^{\prime}}{4\Delta_r} +\frac{1}{2r}
 \pm i\frac{1}{r^2\Delta_r} \Big[(r^2+a^2)(r^2+b^2)\omega -(r^2+b^2)ma \nn \\
&&\qquad\qquad  -(r^2+a^2)kb +Q\big(ab\omega -mb -ka\big)\Big] \, , \nn \\
&& \cL_p^{\pm} = \p_p +\frac{\Delta_p^{\prime}}{4\Delta_p} +\frac{1}{2p} \pm
  \frac{(p^2-a^2)(p^2-b^2)}{p^2\Delta_p}\Big(\omega +\frac{ma}{p^2-a^2}
  +\frac{kb}{p^2-b^2}\Big) \, . \nn
\eea

The separated radial and angular Eqs. (\ref{sdea}-\ref{sded}) can be reduced into a master
equation containing only one component, however, the decoupled master equations are very complicated.
The angular parts can be transformed into the radial part if one replaces $p$ by $ir$ in the vacuum
case where $M = Q = 0$.

In the above, we have explicitly shown that not the usual Dirac equation but a modified one is
separable by variables in the general case of two unequal angular momenta. It is should be pointed
out that all our calculations are done in this general case, the job for equal rotation parameters
$b = \epsilon a$ is a trivial thing, and can be easily completed by using the line element (\ref{Bmab}).
The solution for the separated angular part is a spinorial hyperspherical harmonics, while the
radial parts remain unchanged (just setting $b = \epsilon a$). In fact, the authors of Ref. \cite{DKL}
had already studied the separation of the usual Dirac equation in this special case (with a negative
cosmological). However, they did not consider the effect of the extra counterterm which still makes
an important contribution to the theory. Therefore, their work on the separation of the Dirac equation is
incomplete in this sense. The present work can naturally reduce to theirs in this special case if one
omits the extra counterterm introduced here. In the remaining section, we shall demonstrate that the
existence of a rank-three generalized Killing-Yano tensor is responsible for the separability of the
modified Dirac equation in the five-dimensional rotating, charged Cveti\v{c}-Youm black hole geometry.

%%%%%%%%%%%%%%%%%%%%%%%%%%%%%%%%%%%%%%%%%%%%%%%%%%%%%%%%%
\section{Generalized Killing-Yano tensor and first-order
symmetry operator constructed from it} \label{dualDirac}
%%%%%%%%%%%%%%%%%%%%%%%%%%%%%%%%%%%%%%%%%%%%%%%%%%%%%%%%%

In the last section, we have explicitly shown that the modified Dirac's equation is separable in the
$D = 5$ Cveti\v{c}-Youm black hole spacetime. In this section, we will demonstrate that this separability
is closely related to the existence of a rank-three anti-symmetric tensor admitted by the Cveti\v{c}-Youm
metric. To this end, one should construct a first-order symmetry operator that commutes with the modified
Dirac operator, by using this antisymmetric tensor of rank-three, which should be determined first.

In the case of a five-dimensional uncharged Myers-Perry black hole, it has been demonstrated \cite{KF,WuP1}
that the rank-two St\"{a}ckel-Killing tensor can be constructed from its ``square root,'' a rank-three,
totally antisymmetric Killing-Yano tensor, whose Hodge dual is a rank-two conformal Killing-Yano tensor
that can be generated from a potential one-form.

In order to generalize this research to the rotating, charged Cveti\v{c}-Youm black hole case, one
expects that the rank-two, St\"{a}ckel-Killing tensor given in Eq. (\ref{5dKT}) still can be constructed
from a rank-three antisymmetric tensor,
\be
K_{\mu\nu} = -\frac{1}{2}f_{\mu\alpha\beta}f_{\nu}^{~\alpha\beta} \, .
\ee
Now that in the local Lorentzian coframe (\ref{pentad}), the St\"{a}ckel-Killing tensor has a simple,
diagonal form $K_{AB} = \mbox{diag} (-p^2, p^2, -r^2, -r^2, p^2-r^2)$, it is suggested that the
expected antisymmetric tensor of rank-three can be given by
\be
f = \big(-p~e^0\wedge e^1 +r~e^2\wedge e^3\big)\wedge e^5 \, ,
\label{KYt}
\ee
just like the uncharged case \cite{KF,WuP1}.

Apparently the Hodge dual of the three-form $f$ is a two-form $k = -{^*}f$. Adopting the following
definitions:
\be
k_{\mu\nu} = -({^*}f)_{\mu\nu} = -\frac{1}{6}\sqrt{-g}
 \epsilon_{\mu\nu\alpha\beta\gamma}f^{\alpha\beta\gamma} \, , \qquad
f_{\alpha\beta\gamma} = ({^*}k)_{\alpha\beta\gamma} = \frac{1}{2}
 \sqrt{-g}\epsilon_{\alpha\beta\gamma\mu\nu}k^{\mu\nu} \, ,
\ee
and the convention $\epsilon^{01235} = 1 = -\epsilon_{01235}$ for the totally antisymmetric
tensor density $\epsilon_{ABCDE}$, we find that the two-form is
\be
k = r~e^0\wedge e^1 +p~e^2\wedge e^3 \, ,
\ee
which can be generated from a potential one-form \cite{WuP1}
\be
2\hat{b} = \big(p^2-r^2\big)dt +\frac{(r^2+a^2)(p^2-a^2)a}{b^2-a^2}d\phi
 +\frac{(r^2+b^2)(p^2-b^2)b}{a^2-b^2}d\psi \, .
\ee
Clearly, the two-form $k = -{^*}f$ is closed as usual, $dk = 0$, because $k = d\hat{b}$.

It is worth noting that the above analysis essentially follows the same routine as we did in the
uncharged case \cite{WuP1}. At this stage, one needs to check whether the three-form $f = {^*}k$
is still a rank-three Killing-Yano tensor, and whether its Hodge dual $k = -{^*}f$ is a rank-two,
skew-symmetric, conformal Killing-Yano tensor. However, it is disappointing to find that they no
longer obey, respectively, the ordinary Killing-Yano equation
\be
f_{\alpha\beta\mu;\nu} +f_{\alpha\beta\nu;\mu} \not= 0 \, ,
\ee
and the conformal Killing-Yano equation
\be
k_{\alpha\beta;\gamma} +k_{\alpha\gamma;\beta} -\frac{1}{4}\big(g_{\alpha\beta}k^{\mu}_{~\gamma;\mu}
 +g_{\gamma\alpha}k^{\mu}_{~\beta;\mu} -2g_{\beta\gamma}k^{\mu}_{~\alpha;\mu}\big) \not= 0 \, ,
\ee
or in an equivalent form \cite{PenE}
\be
\mathcal{P}_{\alpha\beta\gamma} \equiv k_{\alpha\beta;\gamma}
 +\frac{1}{4}\big(g_{\beta\gamma}k^{\mu}_{~\alpha;\mu}
 -g_{\gamma\alpha}k^{\mu}_{~\beta;\mu}\big) % \not= 0 \, .
 = \frac{1}{\sqrt{3}}f_{\alpha\beta\mu}F_{\gamma}^{~\mu}
 = \frac{1}{\sqrt{3}}\tilde{F}_{\alpha\beta\mu}k_{\gamma}^{~\mu} \not= 0 \, .
\ee

In the subsequent paper \cite{WutP}, we find that the above rank-three antisymmetric tensor obeys a
modified Killing-Yano equation
\be
f_{\alpha\beta\mu;\nu} +f_{\alpha\beta\nu;\mu}
= \mathcal{W}_{\alpha\beta\mu\nu} +\mathcal{W}_{\alpha\beta\nu\mu} \, ,
\ee
where by construction, we have \cite{WutP}
\be
\mathcal{W}_{\alpha\beta\gamma\lambda} = \frac{1}{2\sqrt{3}}\sqrt{-g}
\epsilon_{\alpha\beta\gamma\mu\nu}f^{\mu\nu\rho}F_{\lambda\rho} \, .
\ee
In the following, this three-form field $f = {^*}k$ will be called a generalized Killing-Yano tensor,
and its Hodge dual two-form $k = -{^*}f$ is a generalized conformal Killing-Yano tensor.

In our previous work \cite{WuP1} done in the uncharged Myers-Perry black hole case, we have adopted the
rank-three Killing-Yano tensor to construct a first-order symmetry operator
\be
\mathbb{H}_f = -\frac{1}{2}\gamma^{\mu}\gamma^{\nu}f^{~~\rho}_{\mu\nu}\nabla_{\rho}
 +\frac{1}{16}\gamma^{\mu}\gamma^{\nu}\gamma^{\rho}\gamma^{\sigma}f_{\mu\nu\rho;\sigma} \, ,
\label{dualop}
\ee
that commutes with the Dirac operator. In the charged Cveti\v{c}-Youm black hole case considered here,
we find that this first-order symmetry operator is precisely the expected one that commutes with the
modified Dirac operator if it is constructed from the generalized Killing-Yano tensor (\ref{KYt}) and
its exterior differential
\be
W = df = -4\Big(\frac{ab}{rp}+\frac{Qp}{r\Sigma}\Big) ~e^0\wedge e^1\wedge e^2\wedge e^3
 -4\sqrt{\frac{\Delta_p}{\Sigma}} ~e^0\wedge e^1\wedge e^2\wedge e^5
 +4\sqrt{\frac{\Delta_r}{\Sigma}} ~e^1\wedge e^2\wedge e^3\wedge e^5 \, .
\ee

At this point, it should be pointed out that the operator $\mathbb{H}_f$ constructed from the above
generalized Killing-Yano tensor obeys the following eigenvalue equation
\be
\big(\mathbb{H}_f +\lambda\big)\Psi = 0 \, ,
\label{dualde}
\ee
in which $\lambda$ is the separation constant introduced in Eqs. (\ref{sdea}-\ref{sded}). Therefore,
separating variables of this equation along the same line as we previously did for the modified
Dirac equation exactly yields the radial and angular parts (\ref{sdea}-\ref{sded}) obtained in the
last section. However, we shall not repeat this process, which is the converse of our construction
procedure below for the operator $\mathbb{H}_f$ in terms of the generalized Killing-Yano tensor.

Our remaining work in this paper is to construct the above first-order symmetry operator commuting
with the modified Dirac operator, parallel to the work \cite{WuP1,WuP2} done in the case of the
five-dimensional Myers-Perry and Kerr-AdS black holes. In what follows, we shall demonstrate that
such a symmetry operator is directly constructed from the separated solutions of the modified
Dirac's equation.

We now proceed to construct such an operator and highlight the construction procedure. According to
our analysis made in the last section, we find that the modified Dirac Eq. (\ref{presde}) can
be split as
\bea
&& \bigg\{ \gamma^0\frac{1}{r^2\sqrt{\Delta_r}}\Big[(r^2+a^2)(r^2+b^2)\p_t +(r^2+b^2)a\p_{\phi}
  +(r^2+a^2)b\p_{\psi} \nn \\
&&\qquad +Q\big(ab\p_t +b\p_{\phi} +a\p_{\psi}\big)\Big] +\gamma^1\sqrt{\Delta_r}\Big(\p_r
 +\frac{\Delta_r^{\prime}}{4\Delta_r} +\frac{1}{2r}\Big) -\frac{i}{r}\big(ab\p_t
 +b\p_{\phi} +a\p_{\psi}\big) \nn \\
&&\qquad\quad +\frac{Q+ab}{2r^2}i\gamma^0\gamma^1 +\mu_er
 -i\lambda\gamma^0\gamma^1\bigg\}\big(\sqrt{r +ip\gamma^5}\Psi\big) = 0 \, , \label{der} \\
&& \bigg\{ \gamma^2\sqrt{\Delta_p}\Big(\p_p +\frac{\Delta_p^{\prime}}{4\Delta_p}
 +\frac{1}{2p}\Big) +\gamma^3\frac{(p^2-a^2)(p^2-b^2)}{p^2\sqrt{\Delta_p}}\Big(\p_t
 -\frac{a}{p^2-a^2}\p_{\phi} -\frac{b}{p^2-b^2}\p_{\psi}\Big) \nn \\
&&\qquad +\frac{\gamma^5}{p}\big(ab\p_t +b\p_{\phi} +a\p_{\psi}\big)
  +\frac{iab}{2p^2}\gamma^0\gamma^1 -i\mu_ep\gamma^5
  +i\lambda\gamma^0\gamma^1 \bigg\}\big(\sqrt{r +ip\gamma^5}\Psi\big) = 0 \, .
\label{dea}
\eea
Now we multiply Eq. (\ref{der}) by $p\gamma^0\gamma^1$ and Eq. (\ref{dea}) by $-r\gamma^2\gamma^3$
respectively from the left and add them together. After using the relations $i\gamma^5 =
\gamma^0\gamma^1\gamma^2\gamma^3$ and $\gamma^2\gamma^3\gamma^5 = i\gamma^0\gamma^1$, we get
a dual equation
\bea
&& \bigg\{ \gamma^0p\sqrt{\Delta_r}\Big(\p_r +\frac{\Delta_r^{\prime}}{4\Delta_r}
 +\frac{1}{2r}\Big) +\gamma^1p\frac{1}{r^2\sqrt{\Delta_r}}\Big[(r^2+a^2)(r^2+b^2)\p_t
 +(r^2+b^2)a\p_{\phi} \nn \\
&&\qquad +(r^2+a^2)b\p_{\psi} +Q\big(ab\p_t +b\p_{\phi} +a\p_{\psi}\big)\Big]
 +\gamma^2(-r)\frac{(p^2-a^2)(p^2-b^2)}{p^2\sqrt{\Delta_p}} \nn \\
&&\qquad \times \Big(\p_t -\frac{a}{p^2-a^2}\p_{\phi} -\frac{b}{p^2-b^2}\p_{\psi}\Big)
 +\gamma^3r\sqrt{\Delta_p}\Big(\p_p +\frac{\Delta_p^{\prime}}{4\Delta_p} +\frac{1}{2p}\Big)
 -i\gamma^0\gamma^1\frac{\Sigma}{rp}\big(ab\p_t \nn \\
&&\qquad +b\p_{\phi} +a\p_{\psi}\big) +\frac{i(Q+ab)p}{2r^2} +\frac{abr}{2p^2}\gamma^5
 +\lambda\big(\gamma^5r -ip\big)\bigg\}\big(\sqrt{r +ip\gamma^5}\Psi\big) = 0 \, .
\eea
Multiplying this equation by the left with a factor $(r+i\gamma^5p)/\Sigma$, we obtain
\bea
&& \bigg\{ \gamma^0p\sqrt{\frac{\Delta_r}{\Sigma}}\Big(\p_r
 +\frac{\Delta_r^{\prime}}{4\Delta_r} +\frac{1}{2r} +\frac{r -ip \gamma^5}{2\Sigma}\Big)
 +\gamma^1p\frac{1}{r^2\sqrt{\Delta_r\Sigma}}\Big[(r^2+a^2)(r^2+b^2)\p_t \nn \\
&&\qquad +(r^2+b^2)a\p_{\phi} +(r^2+a^2)b\p_{\psi} +Q\big(ab\p_t
 +b\p_{\phi} +a\p_{\psi}\big)\Big] \nn \\
&&\qquad +\gamma^2(-r)\frac{(p^2-a^2)(p^2-b^2)}{p^2\sqrt{\Delta_p\Sigma}}\Big(\p_t
 -\frac{a}{p^2-a^2}\p_{\phi} -\frac{b}{p^2-b^2}\p_{\psi}\Big) \nn \\
&&\qquad +\gamma^3r\sqrt{\frac{\Delta_p}{\Sigma}}\Big[\p_p
 +\frac{\Delta_p^{\prime}}{4\Delta_p} +\frac{1}{2p} +\frac{i\gamma^5}{2\Sigma}(r
 -ip\gamma^5)\Big] -i\gamma^0\gamma^1(r+i\gamma^5p)\frac{1}{rp}\big(ab\p_t \nn \\
&&\qquad\quad +b\p_{\phi} +a\p_{\psi}\big) +\frac{iab}{2rp} +\frac{iQp}{2r\Sigma}
 +\gamma^5\Big(\lambda -\frac{ab}{2r^2} +\frac{ab}{2p^2}
 -\frac{Qp^2}{2r^2\Sigma}\Big) \bigg\}\Psi = 0 \, .
\eea
We continue to multiply the above equation a $\gamma^5$ matrix by the left in order to rewrite
it as
\bea
&& \bigg\{ \gamma^5\gamma^0p\sqrt{\frac{\Delta_r}{\Sigma}}\Big(\p_r
 +\frac{\Delta_r^{\prime}}{4\Delta_r} +\frac{1}{2r} +\frac{r -ip \gamma^5}{2\Sigma}\Big)
 +\gamma^5\gamma^1p\frac{1}{r^2\sqrt{\Delta_r\Sigma}}\Big[(r^2+a^2)(r^2+b^2)\p_t \nn \\
&&\qquad +(r^2+b^2)a\p_{\phi} +(r^2+a^2)b\p_{\psi} +Q\big(ab\p_t
 +b\p_{\phi} +a\p_{\psi}\big)\Big] \nn \\
&&\qquad +\gamma^5\gamma^2(-r)\frac{(p^2-a^2)(p^2-b^2)}{p^2\sqrt{\Delta_p\Sigma}}\Big(\p_t
 -\frac{a}{p^2-a^2}\p_{\phi} -\frac{b}{p^2-b^2}\p_{\psi}\Big) \nn \\
&&\qquad +\gamma^5\gamma^3r\sqrt{\frac{\Delta_p}{\Sigma}}\Big(\p_p
 +\frac{\Delta_p^{\prime}}{4\Delta_p} +\frac{1}{2p} +\frac{p+i\gamma^5r}{2\Sigma}\Big)
 +\big(p\gamma^0\gamma^1 -r\gamma^2\gamma^3\big)\frac{1}{rp}\big(ab\p_t \nn \\
&&\qquad\quad +b\p_{\phi} +a\p_{\psi}\big) +\frac{iab}{2rp}\gamma^5
 +\frac{iQp}{2r\Sigma}\gamma^5 +\frac{Q}{2\Sigma} -\frac{Q+ab}{2r^2}
 +\frac{ab}{2p^2} +\lambda \bigg\}\Psi = 0 \, ,
\label{dualeq}
\eea
which can be viewed as an operator eigenvalue equation.

Our last task is to point out that the above Eq. (\ref{dualeq}) is essentially the explicit
form of the eigenvalue Eq. (\ref{dualde}). In order to see this more clearly, one has to find
the explicit expression for the symmetry operator $\mathbb{H}_f$. To construct such an operator
is more involved than to treat with the Dirac operator $\mathbb{H}_D = \gamma^{\mu}\nabla_{\mu}
= \gamma^{\mu}\big(\p_{\mu} +\Gamma_{\mu}\big)$.

Observing the partial differential terms in Eq. (\ref{dualeq}), we can find that it can be exactly
given by $-\frac{1}{2}\gamma^{\mu}\gamma^{\nu}f^{~~\rho}_{\mu\nu}\p_{\rho}$ in terms of the rank-three
generalized Killing-Yano tenor given above, just as the uncharged case. Therefore, we hope to compute
$-\frac{1}{2}\gamma^{\mu}\gamma^{\nu}f^{~~\rho}_{\mu\nu}\nabla_{\rho}$ in the next step. After some
tedious and lengthy algebra manipulations, we get its explicit expression as follows:

\bea
&& -\frac{1}{2}\gamma^{\mu}\gamma^{\nu}f^{~~\rho}_{\mu\nu} \big(\p_{\rho} +\Gamma_{\rho}\big) \nn \\
&&\quad = \gamma^5\gamma^0p\sqrt{\frac{\Delta_r}{\Sigma}}\Big(\p_r
 +\frac{\Delta_r^{\prime}}{4\Delta_r} +\frac{1}{2r} +\frac{r}{2\Sigma}\Big)
 +\gamma^5\gamma^1p\frac{1}{r^2\sqrt{\Delta_r\Sigma}}\Big[(r^2+a^2)(r^2+b^2)\p_t \nn \\
&&\qquad +(r^2+b^2)a\p_{\phi} +(r^2+a^2)b\p_{\psi}
 +Q\big(ab\p_t +b\p_{\phi} +a\p_{\psi}\big)\Big] \nn \\
&&\qquad +\gamma^5\gamma^2(-r)\frac{(p^2-a^2)(p^2-b^2)}{p^2\sqrt{\Delta_p\Sigma}}
 \Big(\p_t -\frac{a}{p^2-a^2}\p_{\phi} -\frac{b}{p^2-b^2}\p_{\psi}\Big)
 +\gamma^5\gamma^3r\sqrt{\frac{\Delta_p}{\Sigma}}\Big(\p_p \nn \\
&&\qquad  +\frac{\Delta_p^{\prime}}{4\Delta_p} +\frac{1}{2p} +\frac{p}{2\Sigma}\Big)
 +\big(p\gamma^0\gamma^1 -r\gamma^2\gamma^3\big)\frac{1}{rp}\big(ab\p_t +b\p_{\phi}
 +a\p_{\psi}\big) +\frac{Q}{2\Sigma} -\frac{Q+ab}{2r^2} \nn \\
&&\qquad\quad +\frac{ab}{2p^2} -\Big(\frac{ab}{rp}+\frac{Qp}{r\Sigma}\Big)i\gamma^5
 +i\sqrt{\frac{\Delta_r}{\Sigma}}\Big(\frac{p^2}{2\Sigma}
 -\frac{3}{2}\Big)\gamma^0 +i\sqrt{\frac{\Delta_p}{\Sigma}}\Big(\frac{3}{2}
 -\frac{r^2}{2\Sigma}\Big)\gamma^3 \, .
\eea

Taking use of the above expression, Eq. (\ref{dualeq}) can be rewritten as
\bea
\bigg\{ -\frac{1}{2}\gamma^{\mu}\gamma^{\nu}f^{~~\rho}_{\mu\nu} \big(\p_{\rho}
 +\Gamma_{\rho}\big) +\lambda +\frac{3i}{2}\gamma^0\sqrt{\frac{\Delta_r}{\Sigma}}
 -\frac{3i}{2}\gamma^3\sqrt{\frac{\Delta_p}{\Sigma}} +\frac{3i}{2}\Big(\frac{ab}{rp}
 +\frac{Qp}{r\Sigma}\Big)\gamma^5 \bigg\}\Psi = 0 \, .
\eea
This equation is almost the one that we expect to seek for the operator $\mathbb{H}_f$ to satisfy
all but the last three terms, which can be supplied by
\be
-\frac{1}{64}\gamma^{\mu}\gamma^{\nu}\gamma^{\rho}\gamma^{\sigma} W_{\mu\nu\rho\sigma}
 = \frac{3i}{2}\bigg[\gamma^0\sqrt{\frac{\Delta_r}{\Sigma}} -\gamma^3\sqrt{\frac{\Delta_p}{\Sigma}}
 +\gamma^5\Big(\frac{ab}{rp}+\frac{Qp}{r\Sigma}\Big)\bigg] \, .
\ee

Therefore we arrive at the final expression for the dual operator $\mathbb{H}_f$ which can be written
as
\be
\mathbb{H}_f = -\frac{1}{2}\gamma^{\mu}\gamma^{\nu}f^{~~\rho}_{\mu\nu}
 \big(\p_{\rho} +\Gamma_{\rho}\big)
 -\frac{1}{64}\gamma^{\mu}\gamma^{\nu}\gamma^{\rho}\gamma^{\sigma} W_{\mu\nu\rho\sigma} \, .
\ee
Using the definition $W_{\mu\nu\rho\sigma} = -f_{\mu\nu\rho;\sigma} +f_{\nu\rho\sigma;\mu}
 -f_{\rho\sigma\mu;\nu} +f_{\sigma\mu\nu;\rho}$ and the identity $f^{\rho}_{~\mu\nu;\rho} = 0$
as well as the property of gamma matrices, one can further recast the above operator into
the equivalent form given by Eq. (\ref{dualop}).

Clearly, the modified Dirac operator $\widetilde{\mathbb{H}}_D$ commutes with its dual operator
$\mathbb{H}_f$, because they obey $\widetilde{\mathbb{H}}_D\Psi = -\mu_e\Psi$ and $\mathbb{H}_f\Psi
= -\lambda\Psi$, respectively. From our discussion in the last section, it is easy to see that
$\lambda$ is a separation constant introduced in the separated radial and angular parts of the
modified Dirac equation. This constant acts as the eigenvalue of a first-order differential
operator $\mathbb{H}_f$ constructed from the ``generalized'' Killing-Yano tensor. In other words,
the dual first-order differential operator characterizes the separation constant introduced in
the separated solutions of the modified Dirac equation and explains why separation of the modified
Dirac equation can be achieved. Therefore, the separability of the modified Dirac equation originates
from the existence of a ``generalized'' Killing-Yano tensor admitted by the three-equal-charge
Cveti\v{c}-Youm metric.

Expanding the commutative relation $[\widetilde{\mathbb{H}}_D, \mathbb{H}_f] = 0$ yields the
generalized Killing-Yano equation and the integrability condition for the generalized Killing-Yano
tensor of rank-three.

To end our discussions, it should be pointed out that the first-order symmetry operator
$\mathbb{H}_f$ can be thought of as the ``square root'' of the second-order operator $\mathbb{K}$.
It has a lot of correspondences in different contexts. It is a five-dimensional charged analogue
to the nonstandard Dirac operator discovered in \cite{CM} for the four-dimensional Kerr metric.
This operator corresponds to the nongeneric supersymmetric generator in pseudoclassical mechanics
\cite{TC,Spcm}.

%%%%%%%%%%%%%%%%%%%%%%%%%%%%
\section{Conclusions}
\label{CoRe}
%%%%%%%%%%%%%%%%%%%%%%%%%%%%

In this paper, we have investigated the separability of a spin-$1/2$ spinor field in the rotating
charged Cveti\v{c}-Youm black hole background spacetime and its relation to a generalized Killing-Yano
tensor of rank-three. Within the f\"{u}nfbein formalism, we have established a suitable pentad for
the Cveti\v{c}-Youm metric and obviously shown that the usual Dirac equation of fermion fields
can not be separated by variables in this general background geometry with two independent angular
momenta. Only when a new additional counterterm is supplemented into the Dirac equation can the
modified Dirac field equation for spin-$1/2$ fermions in the five-dimensional Cveti\v{c}-Youm metric
be decoupled into purely radial and purely angular parts. We have also dealt with the separation
of a massive Klein-Gordon equation in the same background geometry and presented a simple diagonal
form for the St\"{a}ckel-Killing tensor, which can be easily written as the square of a rank-three
generalized Killing-Yano tensor. Two symmetry operators that commute respectively with the scalar
Laplacian operator and the modified Dirac operator have been constructed from the separated solutions
of the massive Klein-Gordon equation and the modified massive Dirac's equation. They have exactly
the same expressions as those obtained in the uncharged Myers-Perry black hole case, and can be
written in terms of the St\"{a}ckel-Killing tensor and a generalized Killing-Yano tensor, respectively.
The success in dealing with the separability of a spin-$1/2$ spinor field equation is due to the
supplement of a suitable counterterm into the usual Dirac equation, the least cost to pay for doing
this is to modify the Killing-Yano tensor equation so that it can be subject to five-dimensional
rotating charged black holes within minimal $D = 5$ EMCS (un-)gauged supergravity theory.

The work presented here includes our previous paper \cite{WuP1} as a special case done for
the general $D = 5$ Myers-Perry metric. In other words, the present work can completely recover all
results \cite{WuP1,WuP2} previously obtained for the uncharged case. In the subsequent paper \cite{WutP},
we will show that the present analysis is directly applicable to deal with the separability of field
equations for spin-$0$ and spin-$1/2$ charged particles in the general, non-extremal, rotating, charged
black holes in minimal $D = 5$ gauged supergravity \cite{CCLP}. It is found that the modified Dirac
equation suggested in this paper can be separated by variables into purely radial and purely angular
parts in this EMCS background spacetime. The Hodge dual of the generalized Killing-Yano tensor of
rank-three is a generalized principal conformal Killing-Yano tensor of rank-two, which can generate
the whole ``tower'' of generalized Killing-Yano and St\"{a}ckel-Killing tensors that are responsible
for the hidden symmetries of this general EMCS-Kerr-AdS black hole geometry. Our research will
further generalize the notion of the principal conformal Killing-Yano tensor that was first introduced
in \cite{KF} (see \cite{VFDK} for a complete summary of its properties) for the higher-dimensional
rotating vacuum black holes.

A possible application of the present work is that starting from our modified Dirac equation, the
gyromagnetic ratio for an ``electron'' can be exactly computed in the five-dimensional three-equal-charge
Cveti\v{c}-Youm black hole spacetime. Another further application is to determine the supersymmetric
condition of this spacetime and confirm that the BMPV black hole is the only supersymmetric, rotating,
charged, asymptotically flat solution in five dimensions. However, these subjects are beyond the scope
of this paper.

\medskip
%%%%%%%%%%%%%%%%%%%%%%%%%%
\textbf{Acknowledgments}:
%%%%%%%%%%%%%%%%%%%%%%%%%%
This work is partially supported by the Natural Science Foundation of China under Grant No. 10675051.

%%%%%%%%%%%%%%%%%%%%%%%%%%%%%%%%%%%%%%%%%%%%%%
\section*{Appendix:~~ Connection one-forms,
curvature two-forms, and other useful tensors}
%%%%%%%%%%%%%%%%%%%%%%%%%%%%%%%%%%%%%%%%%%%%%%

\def\theequation{A\arabic{equation}}
\setcounter{equation}{0}

In this appendix, the spin-connection one-forms, the spinor-connection one-forms, and the curvature
two-forms are presented in the f\"{u}nfbein formalism. Once the pentad (\ref{pentad}) has been
chosen, the exterior differential of the coframe one-forms can be figured out. After some algebraic
computations, we obtain
\bea
&& de^0 = -\Big(\frac{\Delta_r^{\prime}}{2\Delta_r} -\frac{r}{\Sigma}\Big)
 \sqrt{\frac{\Delta_r}{\Sigma}} ~e^0\wedge e^1
 -\frac{p}{\Sigma}\sqrt{\frac{\Delta_p}{\Sigma}} ~e^0\wedge e^2
 -\frac{2p}{\Sigma} \sqrt{\frac{\Delta_r}{\Sigma}} ~e^2\wedge e^3 \, , \nn \\
&& de^1 = -\frac{p}{\Sigma}\sqrt{\frac{\Delta_p}{\Sigma}} ~e^1\wedge e^2 \, , \nn \\
&& de^2 = \frac{r}{\Sigma}\sqrt{\frac{\Delta_r}{\Sigma}} ~e^1\wedge e^2 \, , \nn \\
&& de^3 = \frac{2r}{\Sigma}\sqrt{\frac{\Delta_p}{\Sigma}} ~e^0\wedge e^1
 +\frac{r}{\Sigma}\sqrt{\frac{\Delta_r}{\Sigma}} ~e^1\wedge e^3
 +\Big(\frac{\Delta_p^{\prime}}{2\Delta_p} -\frac{p}{\Sigma}\Big)
 \sqrt{\frac{\Delta_p}{\Sigma}} ~e^2\wedge e^3 \, , \nn \\
&& de^5 = -2\Big(\frac{ab+Q}{r^2p} -\frac{Qr^2}{p\Sigma^2}\Big)  ~e^0\wedge e^1
 +\frac{1}{r}\sqrt{\frac{\Delta_r}{\Sigma}} ~e^1\wedge e^5 \nn \\
&&\qquad\quad +2\Big(\frac{ab}{rp^2} +\frac{Qp^2}{r\Sigma^2}\Big) ~e^2\wedge e^3
 +\frac{1}{p}\sqrt{\frac{\Delta_p}{\Sigma}} ~e^2\wedge e^5 \, .
\eea
The pentad one-forms $e^A$ satisfy the torsion-free condition --- Cartan's first structure Eq.
(\ref{CFE}), through which the spin-connection one-form $\omega^A_{~B} = \omega^A_{~B\mu}dx^{\mu} =
\Upsilon^A_{~BC}e^C$ can be uniquely determined as follows:

\bea
&& \omega^0_{~1} = \Big(\frac{\Delta_r^{\prime}}{2\Delta_r} -\frac{r}{\Sigma}\Big)
 \sqrt{\frac{\Delta_r}{\Sigma}} ~e^0 +\frac{r}{\Sigma}\sqrt{\frac{\Delta_p}{\Sigma}} ~e^3
  -\Big(\frac{ab+Q}{r^2p} -\frac{Qr^2}{p\Sigma^2}\Big) ~e^5 \, , \nn \\
&& \omega^0_{~2} = \frac{p}{\Sigma}\sqrt{\frac{\Delta_p}{\Sigma}} ~e^0
 -\frac{p}{\Sigma}\sqrt{\frac{\Delta_r}{\Sigma}} ~e^3 \, , \nn \\
&& \omega^0_{~3} = \frac{r}{\Sigma}\sqrt{\frac{\Delta_p}{\Sigma}} ~e^1
 +\frac{p}{\Sigma}\sqrt{\frac{\Delta_r}{\Sigma}} ~e^2 \, , \nn \\
&& \omega^0_{~5} =  -\Big(\frac{ab+Q}{r^2p} -\frac{Qr^2}{p\Sigma^2}\Big) ~e^1 \, , \nn \\
&& \omega^1_{~2} = \frac{p}{\Sigma}\sqrt{\frac{\Delta_p}{\Sigma}} ~e^1
 -\frac{r}{\Sigma}\sqrt{\frac{\Delta_r}{\Sigma}} ~e^2 \, , \nn \\
&& \omega^1_{~3} = \frac{r}{\Sigma}\sqrt{\frac{\Delta_p}{\Sigma}} ~e^0
 -\frac{r}{\Sigma}\sqrt{\frac{\Delta_r}{\Sigma}} ~e^3 \, , \nn \\
&& \omega^1_{~5} =  -\Big(\frac{ab+Q}{r^2p} -\frac{Qr^2}{p\Sigma^2}\Big) ~e^0
 -\frac{1}{r}\sqrt{\frac{\Delta_r}{\Sigma}} ~e^5 \, , \nn \\
&& \omega^2_{~3} = -\frac{p}{\Sigma}\sqrt{\frac{\Delta_r}{\Sigma}} ~e^0
 -\Big(\frac{\Delta_p^{\prime}}{2\Delta_p} -\frac{p}{\Sigma}\Big)
 \sqrt{\frac{\Delta_p}{\Sigma}} ~e^3 -\Big(\frac{ab}{rp^2} +\frac{Qp^2}{r\Sigma^2}\Big) ~e^5 \, , \nn \\
&& \omega^2_{~5} = -\Big(\frac{ab}{rp^2} +\frac{Qp^2}{r\Sigma^2}\Big) ~e^3
 -\frac{1}{p}\sqrt{\frac{\Delta_p}{\Sigma}} ~e^5 \, , \nn \\
&& \omega^3_{~5} = \Big(\frac{ab}{rp^2} +\frac{Qp^2}{r\Sigma^2}\Big) ~e^2 \, .
\eea
The local Lorentzian frame component $\Gamma_A$ can be easily read from the spinor-connection one-form
$\Gamma \equiv \Gamma_Ae^A = (1/4)\gamma^A\gamma^B\omega_{AB}$ and is given by
\bea
&& \Gamma_0 = -\Big(\frac{\Delta_r^{\prime}}{4\Delta_r}
 -\frac{r}{2\Sigma}\Big)\sqrt{\frac{\Delta_r}{\Sigma}}\gamma^0\gamma^1
 -\frac{p}{2\Sigma}\sqrt{\frac{\Delta_p}{\Sigma}}\gamma^0\gamma^2
 +\frac{r}{2\Sigma}\sqrt{\frac{\Delta_p}{\Sigma}}\gamma^1\gamma^3 \nn \\
&&\qquad\quad -\Big(\frac{ab+Q}{2r^2p} -\frac{Qr^2}{2p\Sigma^2}\Big)\gamma^1\gamma^5
 -\frac{p}{2\Sigma}\sqrt{\frac{\Delta_r}{\Sigma}}\gamma^2\gamma^3 \, , \nn \\
&& \Gamma_1 = -\frac{r}{2\Sigma}\sqrt{\frac{\Delta_p}{\Sigma}}\gamma^0\gamma^3
 +\Big(\frac{ab+Q}{2r^2p} -\frac{Qr^2}{2p\Sigma^2}\Big)\gamma^0\gamma^5
 +\frac{p}{2\Sigma}\sqrt{\frac{\Delta_p}{\Sigma}}\gamma^1\gamma^2 \, , \nn \\
&& \Gamma_2 = -\frac{p}{2\Sigma}\sqrt{\frac{\Delta_r}{\Sigma}}\gamma^0\gamma^3
 -\frac{r}{2\Sigma}\sqrt{\frac{\Delta_r}{\Sigma}}\gamma^1\gamma^2
 +\Big(\frac{ab}{2rp^2} +\frac{Qp^2}{2r\Sigma^2}\Big)\gamma^3\gamma^5 \, , \nn \\
&& \Gamma_3 = -\frac{r}{2\Sigma}\sqrt{\frac{\Delta_p}{\Sigma}}\gamma^0\gamma^1
 +\frac{p}{2\Sigma}\sqrt{\frac{\Delta_r}{\Sigma}}\gamma^0\gamma^2
 -\frac{r}{2\Sigma}\sqrt{\frac{\Delta_r}{\Sigma}}\gamma^1\gamma^3 \nn \\
&&\qquad\quad -\Big(\frac{\Delta_p^{\prime}}{4\Delta_p} -\frac{p}{2\Sigma}\Big)
 \sqrt{\frac{\Delta_p}{\Sigma}}\gamma^2\gamma^3
 -\Big(\frac{ab}{2rp^2} +\frac{Qp^2}{2r\Sigma^2}\Big)\gamma^2\gamma^5 \, , \nn \\
&& \Gamma_5 = \Big(\frac{ab+Q}{2r^2p} -\frac{Qr^2}{2p\Sigma^2}\Big)\gamma^0\gamma^1
 -\frac{1}{2r}\sqrt{\frac{\Delta_r}{\Sigma}}\gamma^1\gamma^5 \nn \\
&&\qquad\quad -\Big(\frac{ab}{2rp^2} +\frac{Qp^2}{2r\Sigma^2}\Big)\gamma^2\gamma^3
 -\frac{1}{2p}\sqrt{\frac{\Delta_p}{\Sigma}}\gamma^2\gamma^5 \, .
\eea

Taking use of the local Lorentzian frame component $\Gamma_A$ given above and the properties of gamma
matrices together with the relation $\gamma^5 = -i\gamma^0\gamma^1\gamma^2\gamma^3$, we arrive at

\bea
\gamma^A\Gamma_A &=& \gamma^1\sqrt{\frac{\Delta_r}{\Sigma}}\Big(\frac{\Delta_r^{\prime}}{4\Delta_r}
 +\frac{1}{2r} +\frac{r}{2\Sigma} \Big) +\gamma^2\sqrt{\frac{\Delta_p}{\Sigma}}
 \Big(\frac{\Delta_p^{\prime}}{4\Delta_p} +\frac{1}{2p} +\frac{p}{2\Sigma}\Big)
 +\frac{r}{2\Sigma}\sqrt{\frac{\Delta_p}{\Sigma}}\gamma^0\gamma^1\gamma^3 \nn \\
&& -\Big(\frac{ab+Q}{2r^2p} -\frac{Qr^2}{2p\Sigma^2}\Big)\gamma^0\gamma^1\gamma^5
 +\frac{p}{2\Sigma}\sqrt{\frac{\Delta_r}{\Sigma}}\gamma^0\gamma^2\gamma^3
 +\Big(\frac{ab}{2rp^2} +\frac{Qp^2}{2r\Sigma^2}\Big)\gamma^2\gamma^3\gamma^5 \nn \\
 &=& \gamma^1\sqrt{\frac{\Delta_r}{\Sigma}}\Big(\frac{\Delta_r^{\prime}}{4\Delta_r}
 +\frac{1}{2r} +\frac{r}{2\Sigma} \Big) +\gamma^2\sqrt{\frac{\Delta_p}{\Sigma}}
 \Big(\frac{\Delta_p^{\prime}}{4\Delta_p} +\frac{1}{2p} +\frac{p}{2\Sigma}\Big)
 +\frac{r}{2\Sigma}\sqrt{\frac{\Delta_p}{\Sigma}}i\gamma^2\gamma^5 \nn \\
 && -\Big(\frac{ab+Q}{2r^2p} -\frac{Qr^2}{2p\Sigma^2}\Big)\gamma^0\gamma^1\gamma^5
 -\frac{p}{2\Sigma}\sqrt{\frac{\Delta_r}{\Sigma}}i\gamma^1\gamma^5
 +\Big(\frac{ab}{2rp^2} +\frac{Qp^2}{2r\Sigma^2}\Big)i\gamma^0\gamma^1 \nn \\
&=& \gamma^1\sqrt{\frac{\Delta_r}{\Sigma}}\Big(\frac{\Delta_r^{\prime}}{4\Delta_r}
 +\frac{1}{2r} +\frac{r -ip\gamma^5}{2\Sigma}\Big) +\gamma^2\sqrt{\frac{\Delta_p}{\Sigma}}
 \Big(\frac{\Delta_p^{\prime}}{4\Delta_p} +\frac{1}{2p} +\frac{p
 +ir\gamma^5}{2\Sigma}\Big) \nn \\
&& +\Big(\frac{ab}{2r^2p^2} +\frac{Q}{2r^2\Sigma}\Big)i\gamma^0\gamma^1\big(r +ip\gamma^5\big)
  -\frac{Q}{2\Sigma^2}i\gamma^0\gamma^1\big(r -ip\gamma^5\big) \, ,
\eea
where a prime denotes the partial derivative with respect to the coordinates $r$ or $p$.

Using our pentad formalism, the curvature two-forms $\mathcal{R}^A_{~B} = d\omega^A_{~B}
+\omega^A_{~C}\wedge \omega^C_{~B}$ can be concisely expressed by
\bea
&& \mathcal{R}^0_{~1} = \alpha ~e^0\wedge e^1 +2C_1 ~e^1\wedge e^5
 -2C_0 ~e^2\wedge e^3 +2C_2 ~e^2\wedge e^5 \, , \nn \\
&& \mathcal{R}^0_{~2} = \beta ~e^0\wedge e^2 -C_0 ~e^1\wedge e^3
 +C_2 ~e^1\wedge e^5 -C_1 ~e^2\wedge e^5 \, , \nn \\
&& \mathcal{R}^0_{~3} = \beta ~e^0\wedge e^3 -C_3 ~e^0\wedge e^5
 +C_0 ~e^1\wedge e^2 -C_1 ~e^3\wedge e^5 \, , \nn \\
&& \mathcal{R}^0_{~5} = -C_3 ~e^0\wedge e^3 +\gamma ~e^0\wedge e^5
  -C_2 ~e^1\wedge e^2 \, , \nn \\
&& \mathcal{R}^1_{~2} = -C_0 ~e^0\wedge e^3 +C_2 ~e^0\wedge e^5
 +\beta ~e^1\wedge e^2 -C_4 ~e^3\wedge e^5 \, , \nn \\
&& \mathcal{R}^1_{~3} = C_0 ~e^0\wedge e^2 +\beta ~e^1\wedge e^3
 -C_3 ~e^1\wedge e^5 +C_4 ~e^2\wedge e^5 \, , \nn \\
&& \mathcal{R}^1_{~5} = -2C_1 ~e^0\wedge e^1 -C_2 ~e^0\wedge e^2
 -C_3 ~e^1\wedge e^3 +2C_4 ~e^2\wedge e^3 +\gamma ~e^1\wedge e^5 \, , \nn \\
&& \mathcal{R}^2_{~3} = 2C_0 ~e^0\wedge e^1 +2C_4 ~e^1\wedge e^5
 +\delta ~e^2\wedge e^3 +2C_3 ~e^2\wedge e^5 \, , \nn \\
&& \mathcal{R}^2_{~5} = -2C_2 ~e^0\wedge e^1 +C_1 ~e^0\wedge e^2
 +C_4 ~e^1\wedge e^3 +2C_3 ~e^2\wedge e^3 +\varepsilon ~e^2\wedge e^5 \, , \nn \\
&& \mathcal{R}^3_{~5} = C_1 ~e^0\wedge e^3 -C_4 ~e^1\wedge e^2
 +\varepsilon ~e^3\wedge e^5 \, ,
\eea
where
\bea
&& \alpha = \frac{2M(3r^2-p^2)}{\Sigma^3} -\frac{8Qab}{\Sigma^3}
 -\frac{Q^2(10r^2+7p^2)}{\Sigma^4} \, ,  \qquad
 \beta = -\frac{2M(r^2-p^2)}{\Sigma^3} +\frac{2Q^2+4Qab}{\Sigma^3} \, , \nn \\
&& \gamma = -\frac{2M}{\Sigma^2}
 +\frac{Q^2(2r^2+p^2)}{\Sigma^4} \, , \qquad
 \delta = \frac{2M(r^2-3p^2)}{\Sigma^3} -\frac{8Qab}{\Sigma^3}
 -\frac{Q^2(r^2-2p^2)}{\Sigma^4} \, , \nn \\
&& \varepsilon = \frac{2M}{\Sigma^2}
 -\frac{Q^2(r^2+2p^2)}{\Sigma^4} \, , \qquad
 C_0 = \frac{4Mrp}{\Sigma^3}  +\frac{2Qab(r^2-p^2)}{rp\Sigma^3}
 -\frac{Q^2(3r^2+2p^2)p}{r\Sigma^4} \, , \nn \\
&& C_1 = \frac{2Qrp}{\Sigma^3}\sqrt{\frac{\Delta_r}{\Sigma}} \, , \quad
 C_2 = -\frac{2Qr^2}{\Sigma^3}\sqrt{\frac{\Delta_p}{\Sigma}} \, , \quad
 C_3 = -\frac{2Qrp}{\Sigma^3}\sqrt{\frac{\Delta_p}{\Sigma}} \, , \quad
 C_4 = \frac{2Qp^2}{\Sigma^3}\sqrt{\frac{\Delta_r}{\Sigma}} \, . \nn
\eea

Finally, the Ricci tensors, the scalar curvature, and the Einstein tensors for the $D = 5$
Cveti\v{c}-Youm metric are
\bea
&& -R_{00} = R_{11} = -\frac{2Q^2(2r^2+p^2)}{\Sigma^4} \, , \qquad
 R_{22} = R_{33} = \frac{2Q^2(r^2+2p^2)}{\Sigma^4} \, ,
 \qquad R_{55} = \frac{2Q^2(r^2-p^2)}{\Sigma^4} \, , \\
&&\hspace*{5cm}  R = -\frac{2Q^2(r^2-p^2)}{\Sigma^4} \, , \\
&&\qquad -G_{00} = G_{11} = -\frac{3Q^2}{\Sigma^3} \, , \qquad
 G_{22} = G_{33} = \frac{3Q^2}{\Sigma^3} \, , \qquad
 G_{55} = \frac{3Q^2(r^2-p^2)}{\Sigma^4} \, .
\eea

Using the pentad (\ref{pentad}), the $U(1)$ gauge potential one-form can be written as
\be
\mathcal{A} = \frac{\sqrt{3}Q}{2\sqrt{\Delta_r\Sigma}} e^0 \, ,
\ee
the field strength two-form and its corresponding Hodge dual three-form are
\bea
F &=& d\mathcal{A} = \frac{\sqrt{3}Q}{\Sigma^2}\big(r ~e^0\wedge e^1 -p ~e^2\wedge e^3\big) \, , \\
\widetilde{F} &=& {^*}F = \frac{\sqrt{3}Q}{\Sigma^2}\big(p ~e^0\wedge e^1
 +r ~e^2\wedge e^3\big)\wedge e^5 \, .
\eea

The complete Einstein equations are satisfied by the energy-momentum tensor of the
$U(1)$ gauge field
\be
 -T_{00} = T_{11} = -\frac{3Q^2}{2\Sigma^3} \, , \qquad
 T_{22} = T_{33} = \frac{3Q^2}{2\Sigma^3} \, , \qquad
 T_{55} = \frac{3Q^2(r^2-p^2)}{2\Sigma^4} \, .
\ee
Finally, the Maxwell-Chern-Simons equation can be rewritten as
\be
\p_{\nu}\big(\sqrt{-g}F^{\mu\nu}\big) +\frac{1}{2\sqrt{3}}
  \epsilon^{\mu\nu\alpha\beta\gamma}F_{\nu\alpha}F_{\beta\gamma} = 0 \, ,
\ee
and is satisfied by verifying that
\be
d\widetilde{F} = -\frac{4\sqrt{3}Q^2rp}{\Sigma^4}~e^0\wedge e^1\wedge e^2\wedge e^3
 = \frac{2}{\sqrt{3}}F\wedge F \, .
\ee

%%%%%%%%%%%%%%%%%%%%%%%%%%

\end{document}